%% file: Main_PyLMT.tex
\documentclass[fleqn,usenatbib]{mnras}

\usepackage{newtxtext,newtxmath}
\usepackage[T1]{fontenc}

\DeclareRobustCommand{\VAN}[3]{#2}
\let\VANthebibliography\thebibliography
\def\thebibliography{\DeclareRobustCommand{\VAN}[3]{##3}\VANthebibliography}

\usepackage{graphicx}	% Including figure files
\usepackage{amsmath}	% Advanced maths commands
\usepackage{lipsum}
\usepackage{subcaption}
\usepackage{enumitem}
\usepackage{pifont}% http://ctan.org/pkg/pifont\
\usepackage{array}
\usepackage{multirow}
\usepackage{rotating}
\newcolumntype{P}[1]{>{\centering\arraybackslash}p{#1}}
\usepackage{xcolor}

\title[The \texttt{PyLMT} transient detection pipeline]{\texttt{PyLMT}: A transient detection pipeline for the 4-m International Liquid Mirror Telescope}

\author[Pranshu et al.]{Kumar Pranshu,$^{1,2}$\thanks{E-mail: kumarpranshu86@gmail.com}
Kuntal Misra,$^{1}$
Bhavya Ailawadhi,$^{1,3}$
Monalisa Dubey,$^{1,4}$
Naveen Dukiya,$^{1,4}$
Sara Filali,$^{5}$
\newauthor
Paul Hickson,$^{6}$
Brajesh Kumar,$^{7}$
Vibhore Negi,$^{8}$
Jean Surdej$^{1,5}$
\\
$^{1}$Aryabhatta Research Institute of Observational sciencES (ARIES), Manora Peak, Nainital, 263001, India\\
$^{2}$Department of Applied Optics and Photonics, University of Calcutta, Kolkata, 700106, India\\
$^{3}$Department of Physics, Deen Dayal Upadhyaya Gorakhpur University, Gorakhpur, 273009, India\\
$^{4}$Department of Applied Physics, Mahatma Jyotiba Phule Rohilkhand University, Bareilly, 243006, India\\
$^{5}$Institute of Astrophysics and Geophysics, University of Li\`{e}ge, All\'{e}e du 6 Ao$\hat{\rm u}$t 19c, 4000 Li\`{e}ge, Belgium\\
$^{6}$Department of Physics and Astronomy, University of British Columbia, 6224 Agricultural Road, Vancouver, BC V6T 1Z1, Canada\\
$^{7}$South-Western Institute for Astronomy Research, Yunnan University, Kunming, 650500, Yunnan, PR China\\
$^{8}$Inter-University Centre for Astronomy and Astrophysics, Ganeshkhind, Pune, 411007, India
}
\date{Accepted 2025 January 27. Received 2024 December 18; in original form 2024 October 8}

\pubyear{2015}

\begin{document}
\label{firstpage}
\pagerange{\pageref{firstpage}--\pageref{lastpage}}
\maketitle

\begin{abstract}
The International Liquid Mirror Telescope (ILMT) is a 4-m aperture, zenith-pointing telescope with a field-of-view of 22$'$, situated in the foothills of the Himalayas. The telescope operates in continuous survey mode, making it a useful instrument for time-domain astronomy, particularly for detecting transients, variable stars, active galactic nuclei variability, and asteroids. This paper presents the \texttt{PyLMT} transient detection pipeline to detect such transient/varying sources in the ILMT images. The pipeline utilises the image subtraction technique to compare a pair of images from the same field, identifying such sources in subtracted images with the help of convolutional neural networks (CNN) based real/bogus classifiers. The test accuracies determined for the real/bogus classifiers ranged from 94\% to 98\%. The resulting precision of the pipeline calculated over candidate alerts in the ILMT frames is 0.91. It also houses a CNN-aided transient candidate classifier that classifies the transient/variable candidates based on host morphology. The test accuracy of the candidate classifier is 98.6\%. It has the provision to identify catalogued asteroids and other solar system objects using public databases. The median execution time of the pipeline is approximately 29 minutes per image of 17 minutes exposure. Relevant CNNs have been trained on data acquired with the ILMT during the cycle of October--November 2022. Subsequent tests on those images have confirmed the detection of numerous catalogued asteroids, variable stars, and other uncatalogued sources. The pipeline has been operational and has detected 12 extragalactic transients, including 2 new discoveries in the November 2023--May 2024 observation cycle.
\end{abstract}


%%%%%%%%%%%%%%%%%%%%%%%%%%%%%%%%%%%%%%%%%%%%%%%%%%

%%%%%%%%%%%%%%%%% BODY OF PAPER %%%%%%%%%%%%%%%%%%

\section{Introduction}
\label{sec:intro}
Modern sky survey programs have enabled the discovery of transient events in large numbers. Advances in imaging technologies, like the charged coupled device (CCD), along with enhanced computational facilities, have been the key drivers behind the facilities conducting these programs. Furthermore, such surveys have significantly expanded the scope of astronomical research by enabling systematic and continuous monitoring of large portions of the sky. These surveys provide a comprehensive view of the dynamic universe, allowing for the study of a wide variety of celestial objects, from transients occurring in distant galaxies —- such as supernovae, tidal disruption events, and active galactic nuclei variability —- to stellar variability in nearby stars, including pulsating variables and eclipsing binaries. Additionally, they play a crucial role in the detection and monitoring of solar-system objects like asteroids and comets \citep{2019PASP..131g8002Y,2021AJ....161..218D,2013PASP..125..357D}. The ability to capture both rare and common phenomena with such surveys has significantly enhanced our understanding of the life cycle of stars, mechanisms driving cosmic explosion and origins of the building blocks of the solar system. 

The Palomar Transient Factory \citep[PTF;][]{2009PASP..121.1395L} was a fully automated, wide-field survey dedicated to systematic exploration of the optical transient sky. The survey utilised an 8.1 deg$^{2}$ CCD camera mounted on the 48-inch Samuel Oschin telescope at Palomar Observatory to perform imaging. The colours and lightcurves of the detected transients were obtained using the Palomar 60-inch telescope. The Zwicky Transient Facility \citep[ZTF;][]{2019PASP..131a8002B} is a new optical transient survey facility that is the successor to the PTF. It uses the refurbished Palomar 48-inch telescope equipped with a custom-build wide-field CCD camera with a field-of-view (FoV) of 47 deg$^{2}$. The facility yields more than an order of magnitude improvement in survey speed over its predecessor. As a result, it has discovered a plethora of transients over the years. Other survey programs that provide transient alerts include the Panoramic Survey Telescope and Rapid Response System \citep[Pan-STARRS;][]{2016arXiv161205560C}, \textit{Gaia} \citep{2016A&A...595A...1G}, Asteroid Terrestrial-impact Last Alert System \citep[ATLAS;][]{2018PASP..130f4505T}, and Gravitational-wave Optical Transient Observer \citep[GOTO;][]{10.1093/mnras/stac013}.    

Such surveys conduct repeated imaging of the sky, utilising real-time image subtraction to detect new transient events \citep{Cao} and provide alerts for follow-up observations. Large volumes of image data acquired through these surveys in a single night make manual methods of finding transients using this method prohibitive. Hence, several automated techniques for detecting transient candidates using image subtraction have been explored and implemented recently \citep{2007ApJ...665.1246B}. Techniques, like boosted decision trees and random forest classifiers, have been successfully implemented in previous surveys like Nearby supernova factory \citep{2004NewAR..48..637W}, PTF and Intermediate PTF \citep[iPTF;][]{10.1093/mnras/stt1306,2017AAS...22931301B}. Convolutional neural network \citep[CNN;][]{726791} based techniques \citep{2017PASA...34...37A,Cabrera_Vives_2017,10.1093/mnras/stx2161,Duev_2019,Mahabal_2019, 2020MNRAS.497.2641T, refId0, 2023AJ....166..115A, S-PLUS} for transient detection have been introduced in recent programs like the ZTF. Transient identification using shapelet analysis has been explored in \citet{2019AJ....158..172A}. Further categorisation of the candidates into subcategories like SNe, variable stars, asteroids, active galactic nuclei (AGNs), and artefacts based on time-series and context data 
have been implemented for ZTF candidate alerts \citep{Carrasco_Davis_2021, S_nchez_S_ez_2021}. The candidates are confirmed/rejected separately using photometric and spectroscopic follow-up observations.

Another such survey, the 4-m International Liquid Mirror Telescope \citep[ILMT;][]{2006SPIE.6267E..04S,Borra:09, ILMT_Surdej, BKumar2022} has recently begun operations in Devasthal, India. It is the first optical survey telescope in India which performs repeated imaging of the zenith sky with a cadence of 1 day, making it an instrument capable of detecting new and interesting transients \citep{10.1093/mnras/sty298}. As a zenith telescope, it offers several advantages, including low observational air mass, optimal image quality, minimal light pollution, and no time lost to pointing adjustments. This capability of the ILMT motivated the development of a transient detection pipeline called the \texttt{PyLMT}. The objectives of the pipeline are twofold (i) use image subtraction to detect transient candidates automatically and in real-time, and (ii) categorise the detected candidates into distinct classes, streamlining the identification and retrieval of target candidates during the manual vetting process of the pipeline outputs. Additionally, the pipeline makes use of the catalogues namely {\it Gaia} \citep{2016A&A...595A...1G}, the Set of Identifications, Measurements and Bibliography for Astronomical Data \citep[SIMBAD;][]{1991ASSL..171...79E}, and Institute of Celestial Mechanics and Computation of Ephemerides \citep[IMCCE;][]{2006ASPC..351..367B} to confirm or reject candidates based on the type of underlying object (e.g. bright star, known galaxy, known minor planet). The pipeline also uses a context-aware \textit{adaptive detection} approach to reduce false positive detections. A rigorous discussion of all the steps in the pipeline, relevant design parameters, module-wise tests, pipeline implementation, and resulting transient detections and discoveries is presented in this paper.   

The paper is structured as follows. Section~\ref{sec:ILMT} covers the basic structural and functional aspects of the ILMT. The overall description of the pipeline, encompassing the image subtraction module and the transient detection and candidate classification steps, is detailed in Section~\ref{sec:archi}. Various tests made on the pipeline and corresponding results, along with implementation on ILMT science images, are given in Section~\ref{sec:tests}. Finally, conclusions and discussions about the prospects of the pipeline are presented in Section~\ref{sec:concl}.

\section{The 4-m International Liquid Mirror Telescope}
\label{sec:ILMT}

\begin{figure*}
    \centering
    \includegraphics[width=0.8\textwidth]{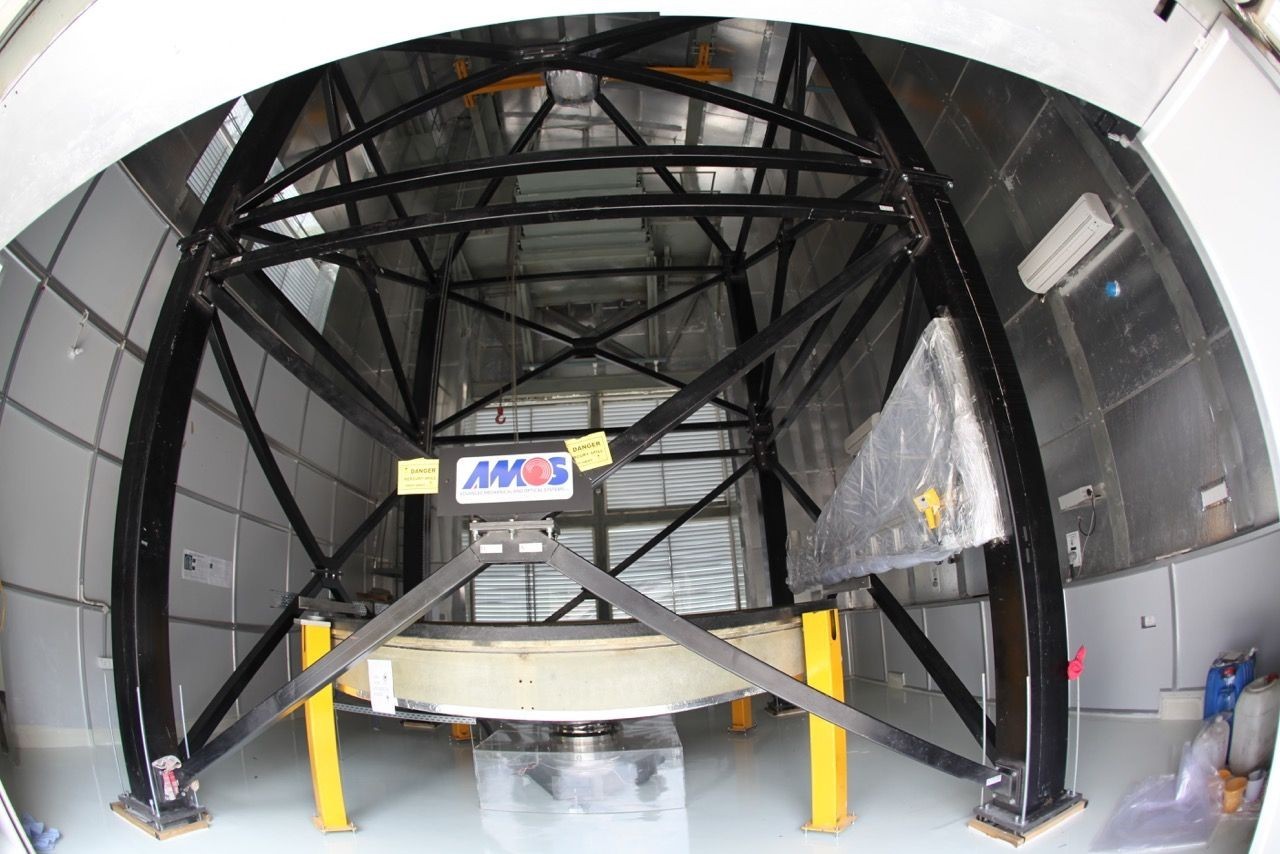}
    \caption{An image of the 4-m ILMT showing the mirror bowl (bottom) mounted on the air bearing, the metallic structure (painted black), four safety pillars (painted yellow), and the CCD-corrector assembly (top).}
    \label{fig:ILMT}
\end{figure*}

The ILMT achieved its first light in April 2022. It is a 4-m aperture \textit{f}/2.36 zenith-pointing telescope with a rotating bowl filled with $\sim$50 litres of mercury constituting the primary reflecting surface. The balance between the centrifugal force and the force of gravity maintains the paraboloid shape of the mirror. The bowl is a carbon fiber-epoxy skin over a closed foam core and is mounted over an air-bearing, supported by a 3-point mount system \citep{2006SPIE.6267E..04S, Borra:09, ILMT_Surdej, BKumar2022}. The mirror is surrounded by a metallic structure that houses the optical corrector and a 4K$\times$4K CCD detector. The telescope's effective FoV is 22$'$. A relatively large aperture enables the detection of faint sources of magnitudes up to 21.9 in SDSS \textit{g}$'$ band \citep{2024BSRSL..93..820A}. The telescope structure is shown in Figure~\ref{fig:ILMT}.

Due to the fixed-pointing nature of the telescope, the time-delay integration (TDI) technique is employed to perform imaging. The technique compensates for the sidereal motion of the sky to give stationary images. The CCD remains fixed and is read out at the exact sidereal rate along the east-west direction. The continuous read-out ensures propagation of accumulated charge at the same sidereal rate, thereby counterbalancing the star trails and rendering point-like stellar images. This technique was first implemented with the Steward Observatory 1.8-m CCD/Transit instrument \citep[CTI;][]{10.1117/12.933448}. It has been observed that the TDI imaging suffers from a north-south elongation at non-zero latitudes due to conical trajectories of stellar images on the CCD plane \citep{1992MNRAS.258..543G}. The optical corrector affixed in front of the focal plane ensures that the conical trajectory traversed by the sources on the CCD plane is made rectilinear \citep{1998PASP..110.1081H,2024BSRSL..93..863N}. The parameters of the ILMT are given in Table~\ref{tab:ILMT_parameters}.

The ILMT covers the same local sidereal time (LST) fields on successive nights. The entire science image covers 1.21 deg$^2$ of the sky and is acquired during a 17-minute exposure in TDI mode. Images are acquired in a single filter for any given night. Depending on the season, up to 35 science frames can be acquired on a typical observation night with favourable weather conditions. These images are processed using the \texttt{PyLMT} pipeline to search for astronomical transients, asteroids, and other variable sources. Up to now, three observation cycles have concluded: October--November 2022, March--June 2023, and November 2023--May 2024.  

\input{ILMT_parameters}

\section{Pipeline Description}
\label{sec:archi}
   
\begin{figure*}
   % \centering
    \includegraphics[width=\textwidth]{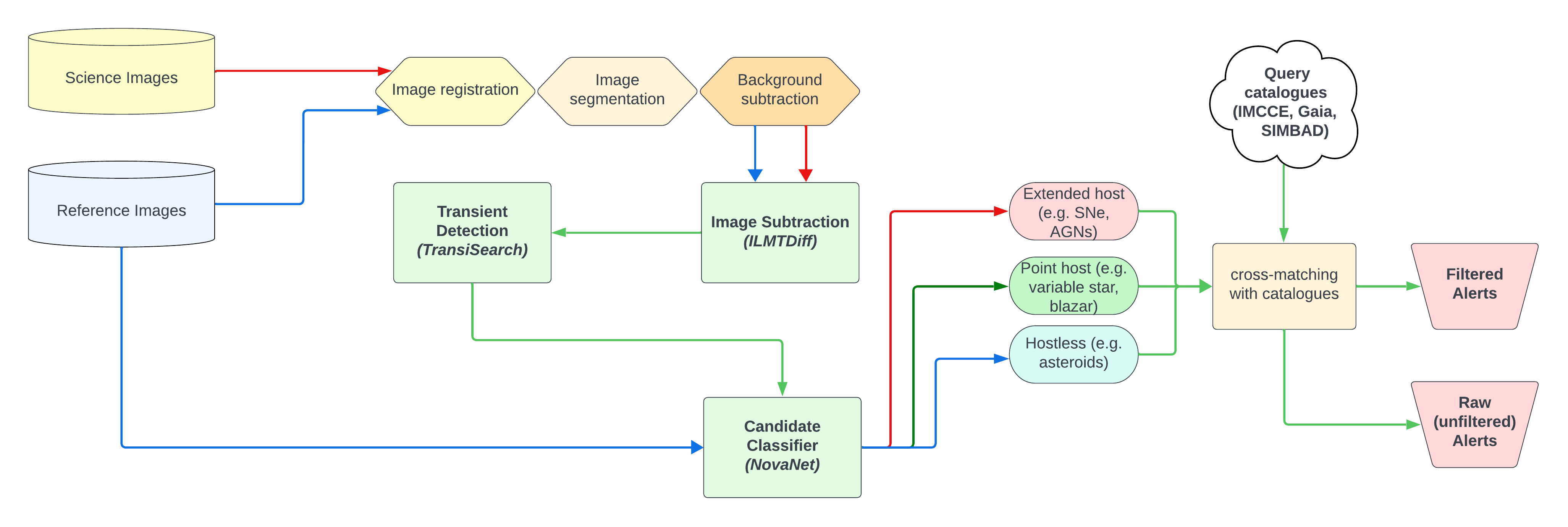}
    \caption{Schematic diagram of the \texttt{PyLMT} transient detection pipeline. The diagram illustrates the three basic steps for carrying out the image subtraction (\texttt{ILMTDiff}), transient detection (\texttt{TransiSearch}), and candidate classification (\texttt{NovaNet}). An additional catalogue crossmatching step identifies catalogued host galaxies and filters out the catalogued minor planets and bright variable stars.}
    \label{fig:PyLMT}
\end{figure*}

The pipeline consists of three modules namely \texttt{ILMTDiff}, \texttt{TransiSearch}, and \texttt{NovaNet} for performing image subtraction, transient detection, and transient candidate classification, respectively. A concise overview of the three basic steps is also provided in \citet{2024BSRSL..93..828P}. Figure~\ref{fig:PyLMT} illustrates the overall architecture of the \texttt{PyLMT} pipeline. Before ingestion to the pipeline, the images have to be corrected for dark, flat, sky background, and astrometry \citep{2022JApA...43...10K,2024BSRSL..93..863N,2024BSRSL..93..837D}. A series of \texttt{Python} and \texttt{Bash} scripts coordinate the various pre-processing and transient detection steps applied to the acquired ILMT images and across multiple workstations. 

\begin{figure*}
    \centering
    \includegraphics[width=\textwidth]{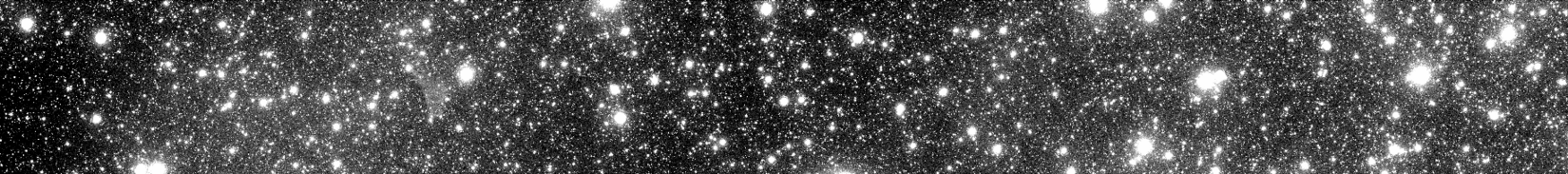}
    \caption{A reduced \textit{i}$'$-band ILMT frame of size 22$'$ along declination and 198$'$ along RA. The first 22$'$ along RA of the acquired full raw frame (not shown here) corresponds to TDI ramping and is therefore sliced out during pre-processing.}
    \label{fig:ILMT_strip}
\end{figure*}

The ILMT frames are acquired as long `strips' with dimensions of 22$'$$\times$220$'$, where the initial 22$'$ corresponds to the TDI ramping region and must be discarded. The TDI ramping occurs because not all sources traverse the full length of the CCD at the start of image acquisition. The east-west orientation of the CCD ensures the orientation of the frames along the RA (refer to Figure~\ref{fig:ILMT_strip}). This results in the frames having unique RA coverage determined by the starting LST and the TDI exposure time (fixed at 17 minutes). 

For every acquired (science) image, a reference image with the nearest acquisition LST is selected from a database. The images are registered using either the WCS information in the \texttt{FITS} header or using the \texttt{astroalign} Python software \citep{beroiz2020astroalign}. An efficient implementation of the image subtraction technique requires a minimal spatial variation of the point spread function (PSF). Therefore, the preprocessed ILMT frames are segmented into 1051$\times$1051 pixel sections before subsequent steps. The adjacent sections overlap by 26 pixels to prevent potential detection failures due to `blind zones' between segments.  

\begin{figure*}
  \centering
  \begin{tabular}{@{}c@{}}
    \includegraphics[width=.33\linewidth]{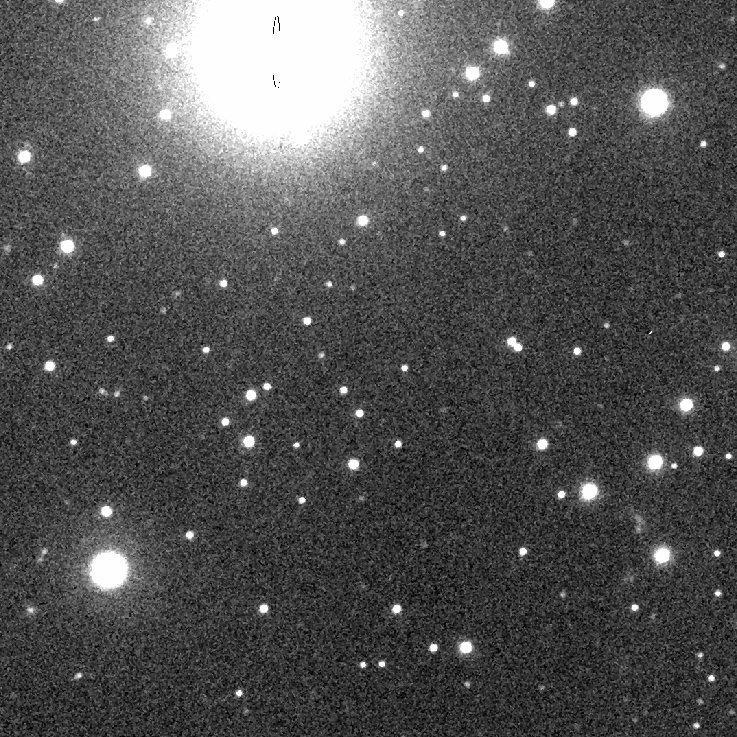} \\[\abovecaptionskip]
    \small (a) Science image
  \end{tabular}
  \hfill
  \begin{tabular}{@{}c@{}}
    \includegraphics[width=.33\linewidth]{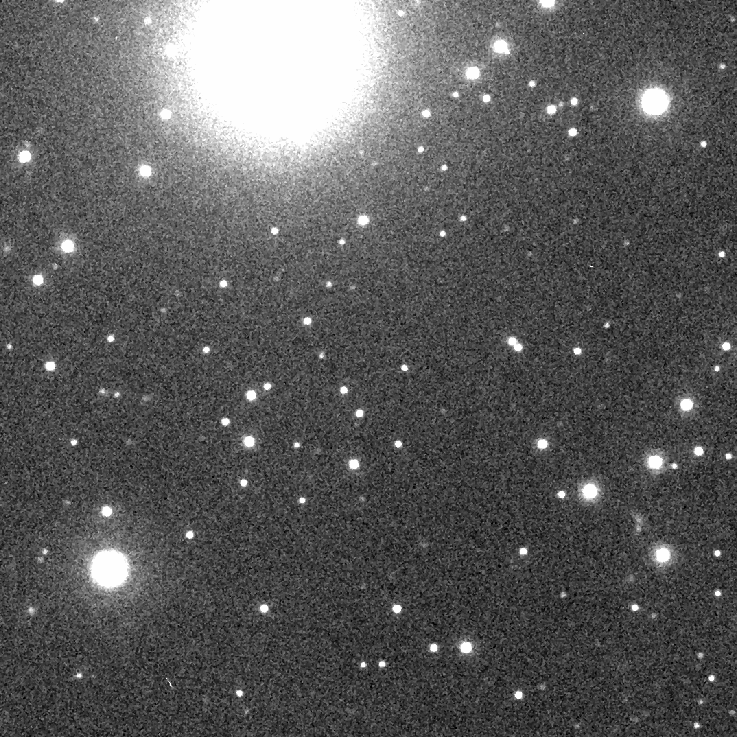} \\[\abovecaptionskip]
    \small (b) Reference image
  \end{tabular}
  \hfill
  \begin{tabular}{@{}c@{}}
    \includegraphics[width=.33\linewidth]{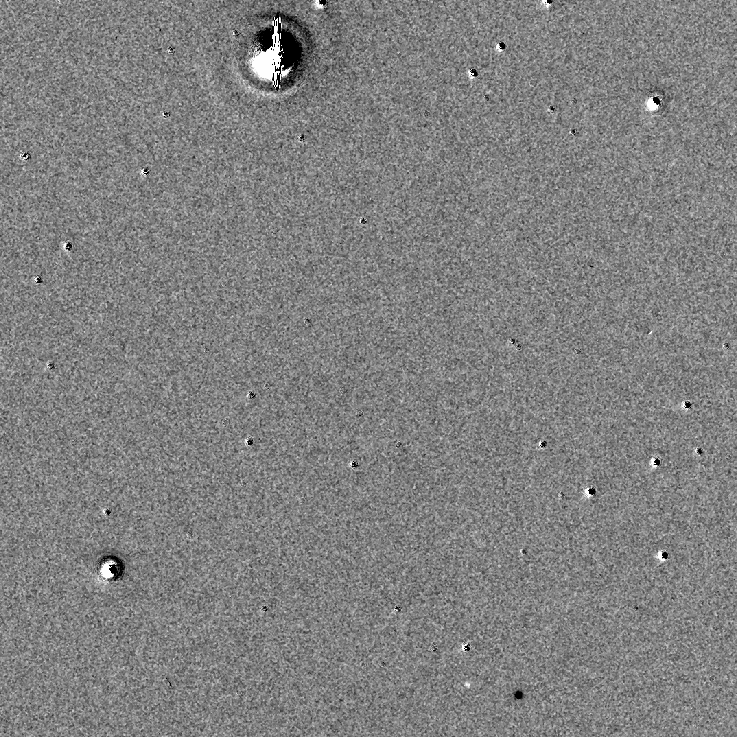} \\[\abovecaptionskip]
    \small (c) Difference image
  \end{tabular}
  \caption{Image subtraction performed on a 1051$\times$1051 pixel ($\sim5'.66 \times 5'.66$) cutout image of an \textit{r}$'$-band ILMT frame using the difference imaging algorithm (\texttt{ILMTDiff}).}
\label{fig:subtraction}
\end{figure*}

\subsection{Image Subtraction}
\label{sec:sub}
Images from survey telescopes like the ILMT are typically populated with numerous astronomical sources. The sources exhibiting non-varying flux are removed to facilitate the isolation of transient and variable sources. This is accomplished using the image subtraction technique wherein a reference image - representing the same region of the sky - is subtracted from the science image to eliminate non-varying sources. The \texttt{ILMTDiff} module performs the image subtraction in the pipeline.  

The science and reference images are first registered and then transferred to this module. The two important steps performed within the module are flux scaling and PSF matching. The science and reference images are acquired on two different nights with possibly different visibility and seeing conditions. This results in image pairs having different source full width at half maximum (FWHM) and flux counts. Image subtraction requires these two parameters for both images (cutouts in our case) to be matched. Scaling for flux counts is performed by multiplying the reference cutout image with a factor equal to the ratio of source fluxes in science and reference images. The reference image is then convolved with an optimal convolution kernel to match the PSFs. This ideally requires the reference image to be acquired in better-seeing conditions (hence smaller PSF FWHM) than the science image as convolution degrades the PSF \citep{10.1093/mnras/stu835}. Finally, the science and convolved reference images are subtracted to produce a difference image. Figure~\ref{fig:subtraction} illustrates the result of image subtraction performed on a 1051$\times$1051 pixel cutout of an ILMT image.

This work builds upon \citet{Bramich_2008}, where the convolution kernel is represented as a square matrix with pixel entries being the parameters to be optimised. Image subtraction is transformed into an optimisation problem, with the squared sum of subtraction residuals being the objective function to be minimised. Residuals are evaluated for pairs of common individual sources in science and reference cutouts. A \texttt{scipy} \citep{Virtanen_2020} hosted numerical least-squares optimiser is used to determine the optimal kernel. Detailed discussion on the \texttt{ILMTDiff} module is given in Appendix~\ref{app:Image_subtraction}.

\begin{figure}
    %\centering
    \includegraphics[width=\columnwidth]{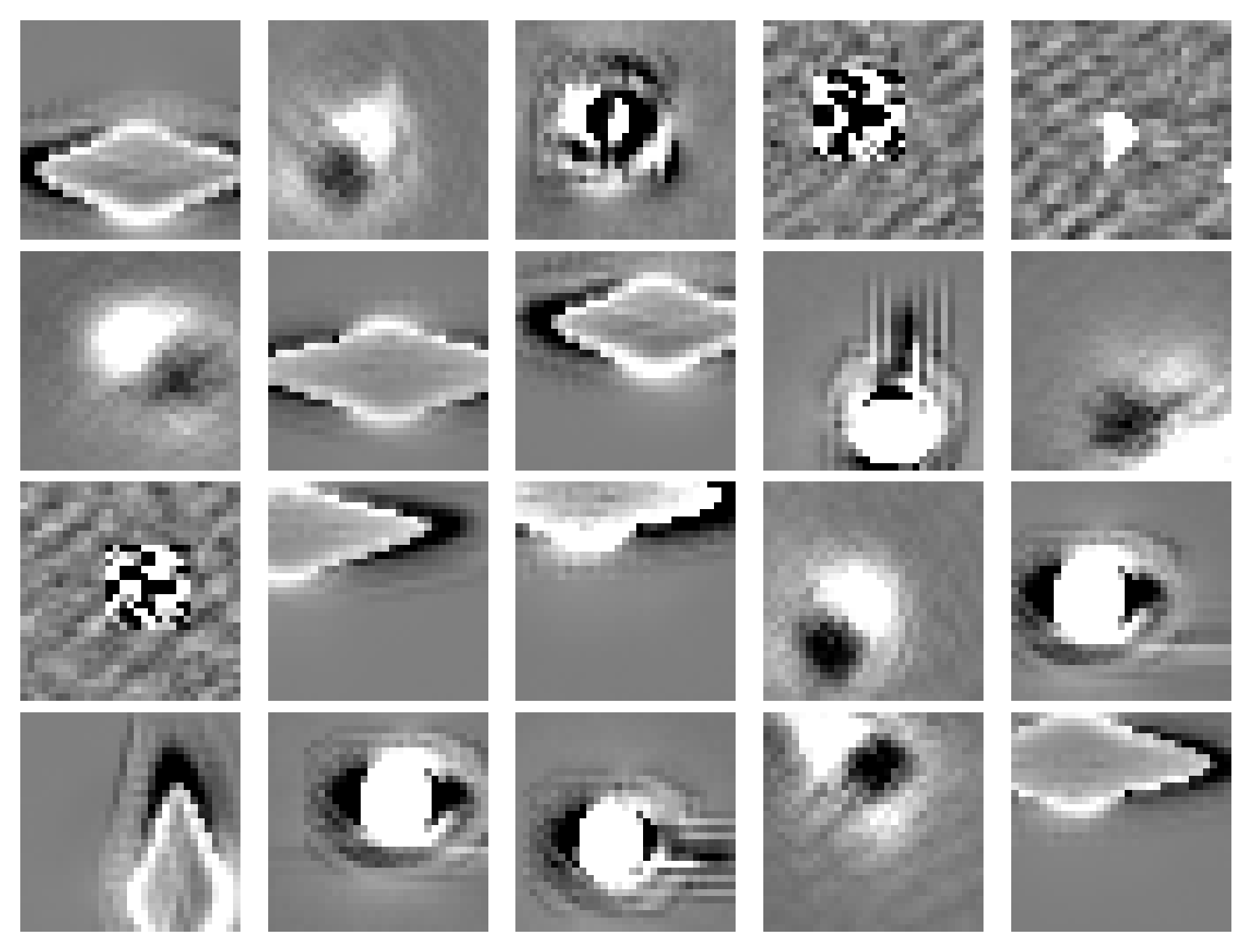}
    \caption{Samples from the bogus dataset for the \textit{real/bogus} dataset. Such artefacts occurring in subtracted images generally result from image misalignment due to poor astrometric calibration, PSF mismatch during subtraction, saturated stars, CCD artefacts and cosmic hits. Each cutout is of size 31$\times$31 pixels ($\sim10'' \times 10''$).}
    \label{fig:bogus}
\end{figure} 

\begin{figure}
    %\centering
    \includegraphics[width=\columnwidth]{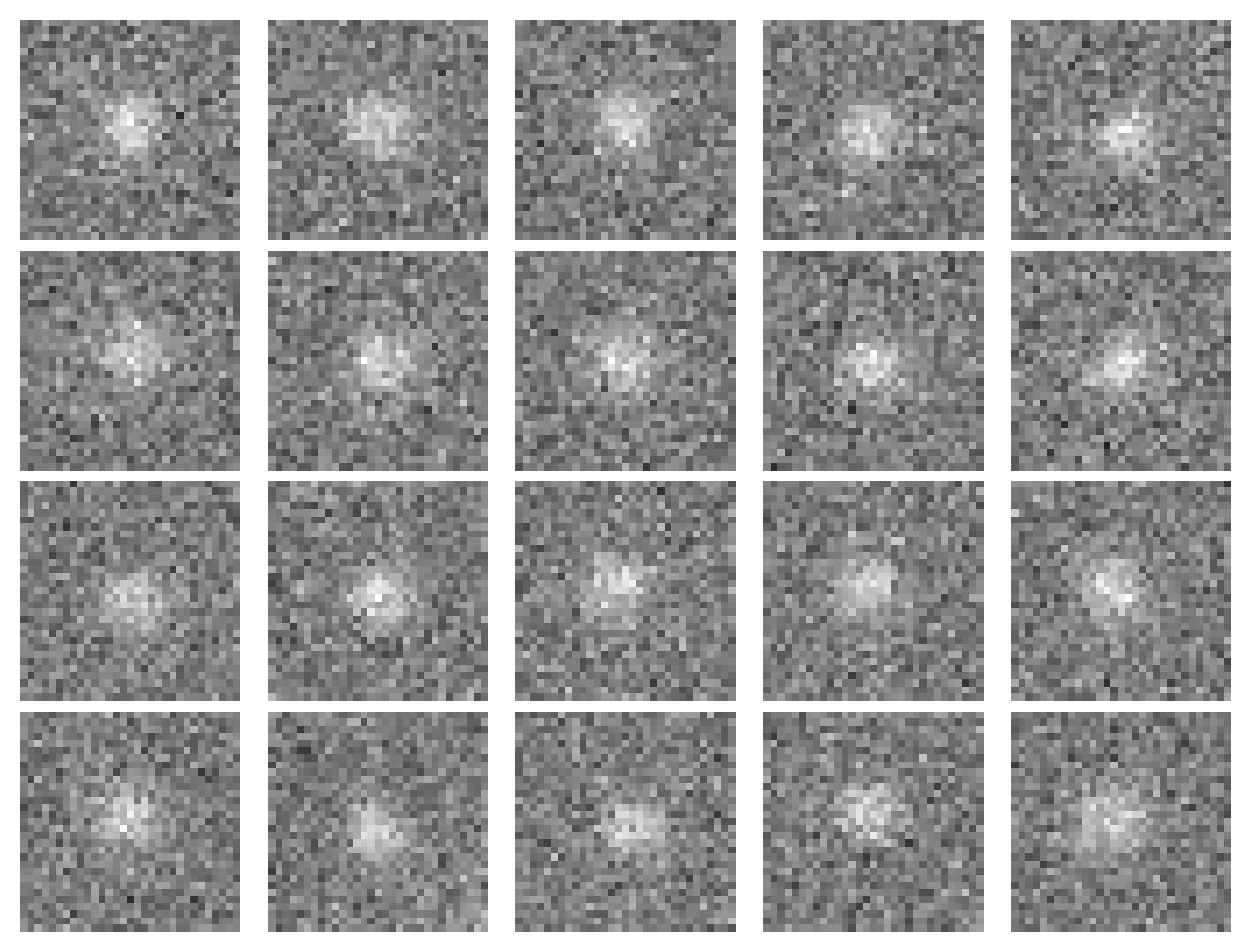}
    \caption{Samples from the real dataset for the \textit{real/bogus} dataset. Gaussian noise was added while preparing the samples to enhance model sensitivity. Each cutout is of size 31$\times$31 pixels ($\sim10'' \times 10''$).}
    \label{fig:real}
\end{figure}

\subsection{Transient Detection}
\label{sec:detect}

\input{transisearch_threshold}

Detecting transients in subtracted images requires filtering of the artefacts that can obscure genuine detection. Common sources of these artefacts include astrometric misalignment, improper subtraction, cosmic hits, random background fluctuation, and defective CCD columns. They are typically characterised by their `non-PSF-like' morphology (refer to Figure \ref{fig:bogus}), which deviates from the expected profile of `real' astrophysical sources. A two-step strategy was adopted for the \texttt{TransiSearch} transient detection module to remove the artefacts. In the first step, the CNN-based `real/bogus' classifier is implemented to reject most artefacts (a.k.a `bogus' sources). Subsequently, threshold-based filtering is applied to the classifier output to enhance the purity of the candidates. Table~\ref{tab:transisearch_threshold} lists the source parameters and respective thresholds used in the module.

A training dataset of nearly 5000 samples was created to train the CNN classifier. The `real' training set consisted of `PSF-like' real astrophysical sources (Figure \ref{fig:real}) extracted from the ILMT frames. The `bogus' training set consisted of visually inspected artefact sources extracted from the subtracted images. Every sample in the dataset has a dimension of 31$\times$31 pixels and is the same as the input dimension of the CNN model. To reduce false positive detections, a two-CNN-based \textit{adaptive detection} strategy is used (discussed in detail in Section \ref{sec:adaptive_detection}).

The \texttt{TransiSearch} module extracts all the sources above 4$\sigma$ significance in the subtracted image using \texttt{find\_peaks} algorithm of the \texttt{Photutils} python software \citep{2016ascl.soft09011B}. The sources are extracted as square cutouts of the dimension 31$\times$31 pixels. They are subsequently preprocessed and classified using the trained CNN model discussed before. The CNN assigns a score to each of the detected sources. A classification threshold of 0.5 is applied to these scores to classify respective sources as `real' or `bogus'. Dataset preparation and training of the `real/bogus' classifier is discussed in detail in Appendix~\ref{app:Real/Bogus}.     

\subsection{Transient Candidate Classification}
\label{sec:class}

\input{candidate_classifier_dataset}
  
Multiple transient candidates are generated in the previous step for every ILMT frame. Real-time classification of the candidates is pivotal to planning the most appropriate follow-up observations. In particular, the early-time photometric and spectroscopic follow-up of SNe can be advantageous in obtaining constraints on its physical properties \citep{Khazov_2016}. The \texttt{NovaNet} module in the pipeline classifies the \texttt{TransiSearch} candidates into three categories namely `extended-host', `point-host' and `hostless'. Following the classification, the candidates undergo cross-matching with existing catalogues (discussed in Section \ref{subsec:catalog_xmatch}), thereby streamlining their selection process for follow-up observations. 

The classification is based on the morphology of the object (if present) in the reference image around the position of detection. The `extended-host' candidates are characterised by an associated extended galaxy (e.g. SN with an extended host galaxy, AGN), `point-host' by a star/point-like source (e.g. variable star, QSO), and `hostless' by the absence of source in the reference image (e.g. asteroids, CVs, SN with faint/invisible host). The `extended-host' candidates, due to their potential association with bright SNe, are prioritised as triggers for follow-up observations.

\begin{figure}
    \centering
    \includegraphics[width=0.47\textwidth]{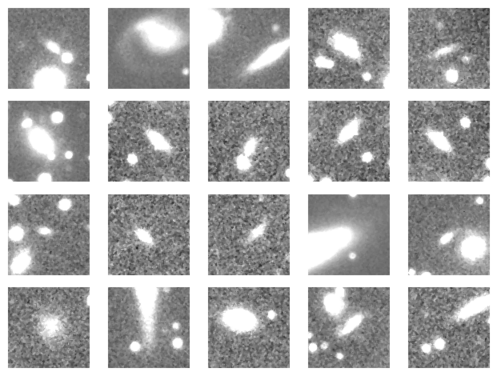}
    \caption{Samples from extended-host (galaxy image) training data for the \texttt{NovaNet} transient candidate classifier. Each cutout is of size 102$\times$102 pixels ($\sim33'' \times 33''$).}
    \label{fig:transi_galaxy}
\end{figure}

\begin{figure}
    \centering
    \includegraphics[width=0.47\textwidth]{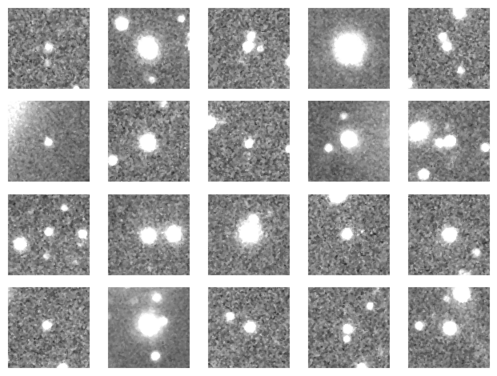}
    \caption{Samples of point-source training data for the \texttt{NovaNet} transient candidate classifier. Each cutout is of size 102$\times$102 pixels ($\sim33'' \times 33''$).}
    \label{fig:transi_stellar}
\end{figure}

\begin{figure}
    \centering
    \includegraphics[width=0.47\textwidth]{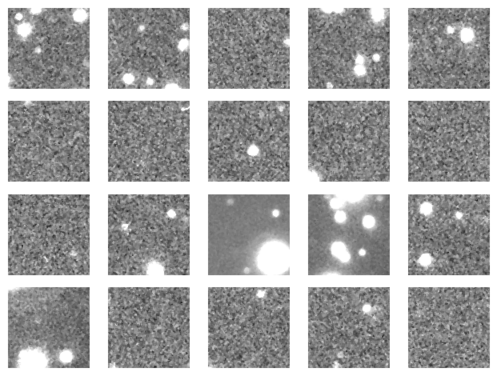}
    \caption{Samples for hostless class training data for the \texttt{NovaNet} transient candidate classifier. Each cutout is of size 102$\times$102 pixels ($\sim33'' \times 33''$).}
    \label{fig:transi_hostless}
\end{figure}

The module uses a combination of trained CNN-based transient candidate classifiers to perform candidate classifications. The training dataset for the CNNs was compiled with around 1600 cutout samples curated with 102$\times$102 pixel image cutouts of galaxies, stars, and empty spaces (Figures~\ref{fig:transi_galaxy}, \ref{fig:transi_stellar} and \ref{fig:transi_hostless}). Data augmentation techniques were employed to enhance the strength and diversity of the dataset. Additionally, a separate dataset with a cutout size of 31$\times$31 pixel was compiled to train another transient candidate classifier for that specific input size. This separate dataset was constructed by cropping the central 31$\times$31 pixel portion from 102$\times$102 pixel samples in the original classifier dataset. The training datasets for candidate classifiers are summarised in Table \ref{table:sample_counts_tr}. When a source is detected at any position in the difference image, a 102$\times$102 pixel cutout and a 31$\times$31 pixel cutout are extracted from the reference image around that position. The two cutouts are then pre-processed and passed through an \textit{ensemble} of multiple CNNs for classification. The final classification of the candidate is determined based on CNN scores associated with these two cutouts. A detailed discussion on dataset preparation, training, and logical organisation of the CNNs is given in Appendix~\ref{app:transient_classifier}.

\subsection{Catalogue cross-matching} \label{subsec:catalog_xmatch}

Early-time identification of transient candidates like SNe is the primary goal of the \texttt{PyLMT} pipeline. After the majority of the artefacts have been removed in the previous steps, astrophysical sources like asteroids and variable stars turn out to be the major contaminants that should be removed. To that end, a catalogue cross-matching step was integrated into the pipeline. This additional step generates two distinct sets of candidates for each frame: (1) A comprehensive list consisting of all the detected transient and variable sources, and (2) a filtered list consisting of candidates after rejecting catalogued asteroids and probable variable stars.

The cross-matching step is hosted within the \texttt{NovaNet} module. The query to respective catalogues is performed using \texttt{astroquery} \citep{2019AJ....157...98G} in \texttt{Python}. The \texttt{SkyBot} service \citep{2006ASPC..351..367B} of the IMCCE is queried for all the candidates to reject solar-system objects within 10$''$ from their detection position. Subsequently, the \textit{Gaia} G-band magnitude is queried for the remaining `point-host' candidates within 1$''$ of the detection position. Such candidates with a corresponding magnitude brighter than 19 are rejected for being probable variable stars. The magnitude threshold is set to ensure that any distant SNe candidate, whose host galaxy might appear as a faint point source, is not erroneously rejected.

\begin{figure*}
    \centering
    \includegraphics[width=\textwidth]{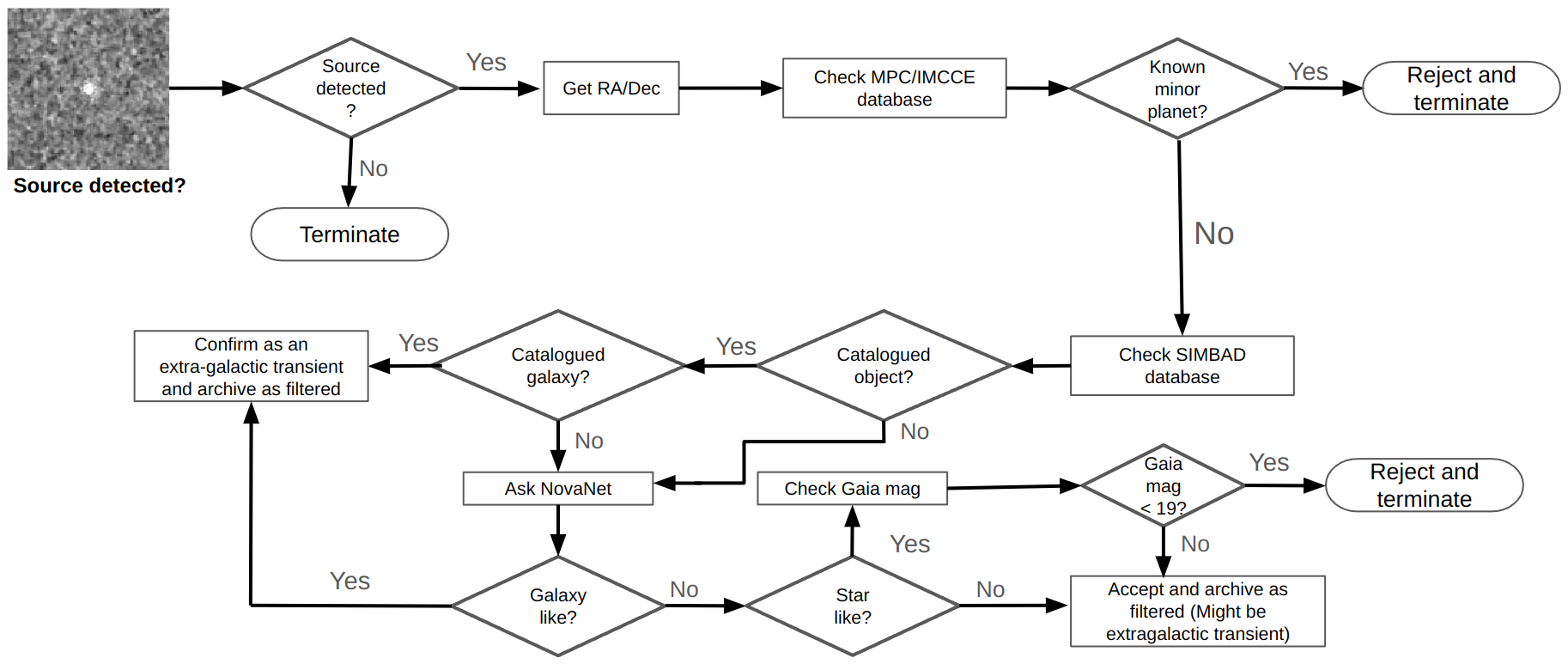}
    \caption{Flowchart illustrating candidate filtering performed using the catalogue cross-matching step in the pipeline.}
    \label{fig:cat_xmatch8}
\end{figure*}

The SIMBAD catalogue is queried to identify any known object (e.g. a catalogued galaxy, star, QSO, etc) underlying the detected candidates within 10$''$ from the position of detection. For every matched counterpart, the object name and type are recorded under host-name entry in the transient candidate alert (refer to Section \ref{sec:command_line}). This is particularly useful in indicating the detection of variable stars, QSOs, CVs, or SNe with established host galaxies (see Figures \ref{tab:detections_trans_1} and \ref{tab:detections_var}). Candidates associated with a catalogued galaxy are retained in the filtered outputs; for other cases, the SIMBAD cross-matching step serves as a supplementary feature without altering the final classification status. Extended-host candidates that do not correspond to known solar-system objects are prioritised for follow-up observations and reported to the transient name server (TNS). The flowchart representation of the entire catalogue cross-matching step is shown in Figure~\ref{fig:cat_xmatch8}.  

\subsection{Adaptive Detection} 
\label{sec:adaptive_detection}

\input{real_bogus_dataset}

`Precision' and `recall' are two important metrics used to evaluate the performance of binary classifiers like the `real/bogus' classifier. They can be expressed mathematically using the following relations:
\[
\text{Recall} = \frac{\text{True Positives}}{\text{True Positives + False Negatives}}
\]
\[
\text{Precision} = \frac{\text{True Positives}}{\text{True Positives + False Positives}}
\]
A model optimised for high recall is capable of detecting fainter transients with a lower signal-to-noise (S/N) ratio, simultaneously being more susceptible to false positive outputs. Alternatively, a model optimised for higher precision exhibits a reduced false positive rate but at the cost of an increased fraction of missed detections, especially at fainter magnitude limits. A solution consisting of context-aware implementation of both varieties of models was adopted to minimise false positives while maintaining acceptable levels of detection sensitivity. 

Upon performing a simple inspection of the candidate output, it was established that the `point-host' candidates (primarily due to improper subtraction of non-variable stars) represented the majority of false positive detections. The adopted solution involves a preemptive classification of all sources (both real and bogus) in the difference image into `point-host' and `non-point-host' categories using CNN. Sources classified as `point-host' by the CNN are forwarded to a high-precision classifier and the rest to a high-recall classifier. This ensures that `point-host' sources, which previously contributed to a significant number of false positives, undergo more rigorous scrutiny compared to other sources. This provision prioritises the detection of fainter, high-priority `non-point-host' transients (e.g. faint SNe), while extremely faint `point-host' transients or variable stars may remain undetected. This provision was implemented within the \texttt{TransiSearch} module itself.

The original `real/bogus' classifier can be used as the high-recall classifier while a separate classifier was needed for the high-precision classifier. One possible method to create such a classifier was to take a high-recall CNN model and increase the classification threshold. Another method was to bias the training dataset towards artefact samples. This makes the trained model better at identifying and rejecting artefacts, thereby reducing false positives. The second method was adopted for the \texttt{NovaNet} module. The implementation of this technique resulted in a reduction of false positive detections in final output from around 35$\%$ to around 10$\%$. The training datasets for high-precision and high-recall classifiers are summarised in Table \ref{table:sample_counts_rb}.

\subsection{Command line implementation and candidate alert interpretation}
\label{sec:command_line}
The \texttt{PyLMT} is readily executable using \texttt{Linux} command line interface (CLI) from inside the science image directory. A list of tunable parameters of the pipeline is listed in Table~\ref{tab:PyLMT_parameters}. The path to the reference image directory has to be specified with the command. The appropriate reference image is selected for a science image using the astrometric information stored in the \texttt{FITS} header. The result for transient detection and classification (referred to as alerts) is stored in a \texttt{zip} folder containing relevant \texttt{PDF} and \texttt{CSV} files.

\begin{figure*}
\setlength{\arrayrulewidth}{0.5mm}
    \centering
    \begin{tabular}{|c|c|}
        \hline
         \includegraphics[width=0.48\linewidth]{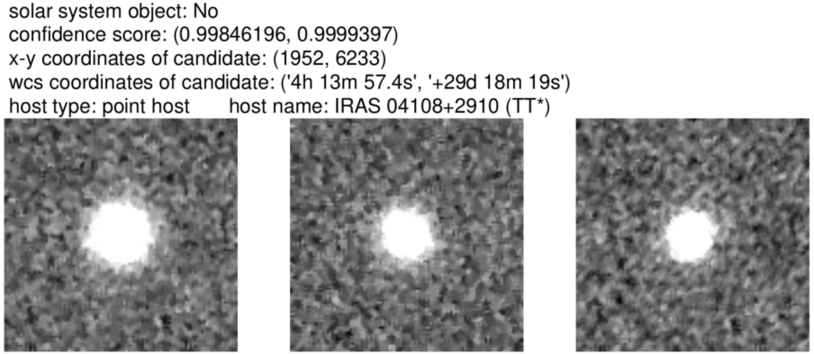} & \includegraphics[width=0.48\linewidth]{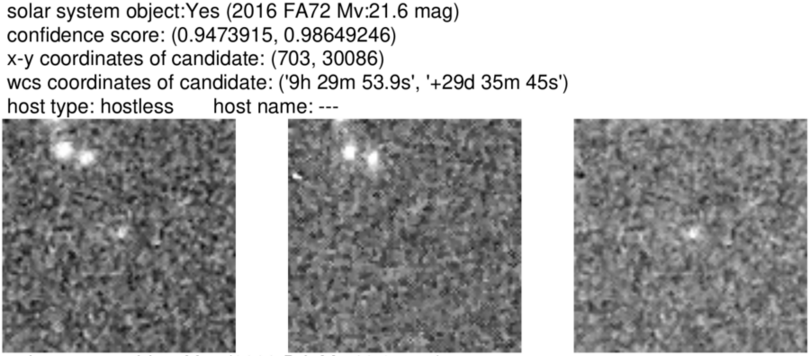}\\
        \hline
        \includegraphics[width=0.48\linewidth]{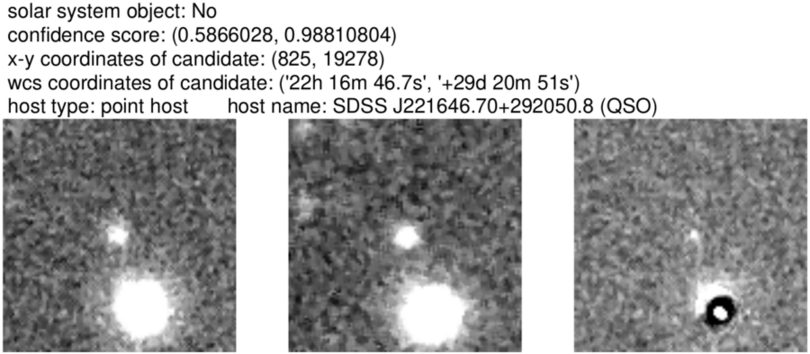} & \includegraphics[width=0.48\linewidth]{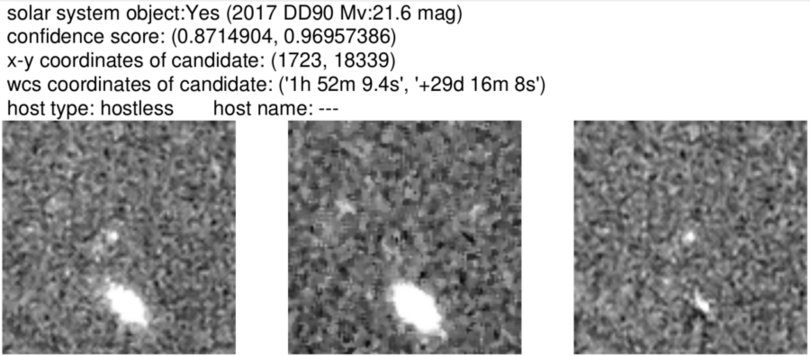} \\
        \hline
        \includegraphics[width=0.48\linewidth]{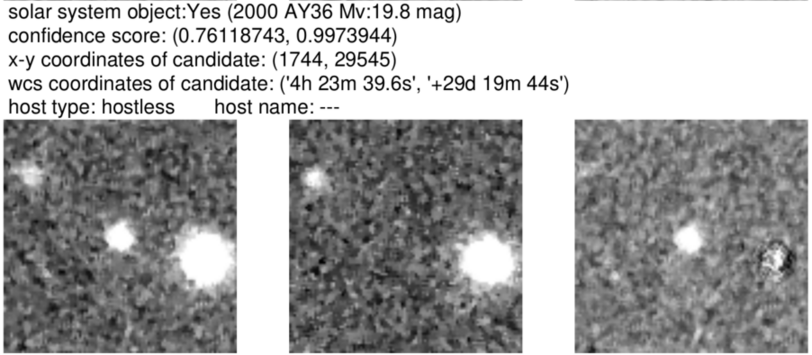} & \includegraphics[width=0.48\linewidth]{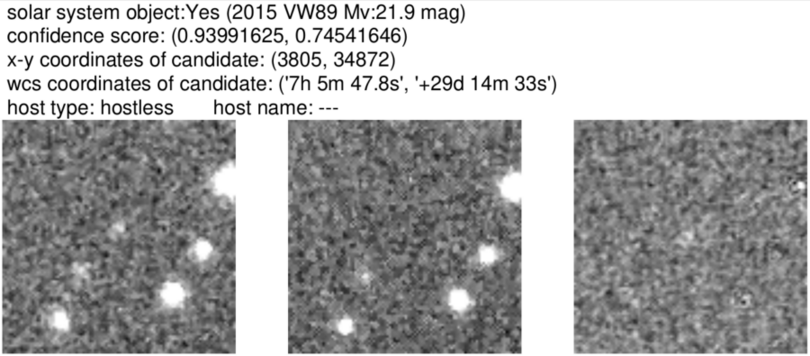} \\
        \hline
    \end{tabular}
    \caption{\textbf{Left column --} From top to bottom: Detection of variability in the known T-Tauri star IRAS 04108+2910, the known QSO SDSS\_J221646.70+292050.8 and the catalogued asteroid 2000 AY36. All these sources were rejected from the filtered list of candidates but were included in the unfiltered list of candidates. If a variable source like a variable star or a QSO is cross-matched with SIMBAD then its name appears as a \textit{`host name'} entry while any catalogued asteroid/minor planet appears as \textit{`solar system object'} entry in the alert. \textbf{Right column --} From top to bottom: Examples of faint asteroids identified using the pipeline, with V-band magnitudes exceeding 21.5 as reported by the MPC.}
    \label{tab:detections_var}
\end{figure*}

Each candidate alert contains six entries (shown in Figures~\ref{tab:detections_trans_1} and \ref{tab:detections_var}), namely \textit{solar system object} which indicates whether it is a minor planet (if yes then the corresponding MPC name and V-band magnitude is printed), \textit{confidence score} which is an ordered pair of CNN confidence scores of classification and detection respectively, \textit{X-Y coordinates} and \textit{WCS coordinates} of candidate, \textit{host type} which indicates the corresponding \texttt{NovaNet} classification and \textit{host name} which indicates the source name retrieved from \texttt{SIMBAD} corresponding to the candidate (e.g. a catalogued host galaxy, variable star, QSO, etc.). The date and starting UTC of the processed science image is also mentioned in the final output. 

\section{Pipeline validation and Results}
\label{sec:tests}
The pipeline was validated on ILMT images to ensure a seamless real-time execution when integrated with the incoming data. The execution time of the pipeline depends on factors like the image quality and crowdedness of the field. The median execution time per ILMT frame (dimensions 4K$\times$36K) is approximately 29 minutes when computed using the Intel\textsuperscript{\tiny\textregistered} Xeon\textsuperscript{\tiny\textregistered} Silver 4210R 10-core CPU clocking at 2.40 GHz. With an estimated acquisition of 35 frames per night, the total processing time for a complete night's dataset is nearly 17 hours. The performance of the three main steps of the pipeline, viz. image subtraction, transient detection, and transient candidate classification, were evaluated using individual tests.

\begin{figure*}
  \centering
  \begin{tabular}{@{}c@{}}
    \includegraphics[width=.24\linewidth]{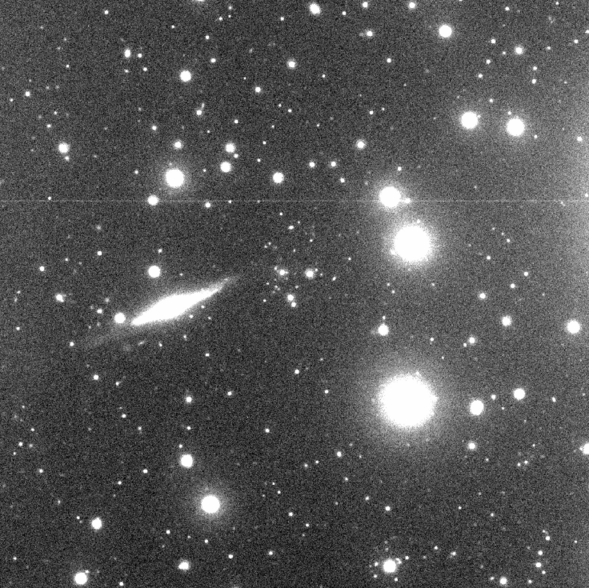} \\[\abovecaptionskip]
  \end{tabular}
  \hfill
  \begin{tabular}{@{}c@{}}
    \includegraphics[width=.24\linewidth]{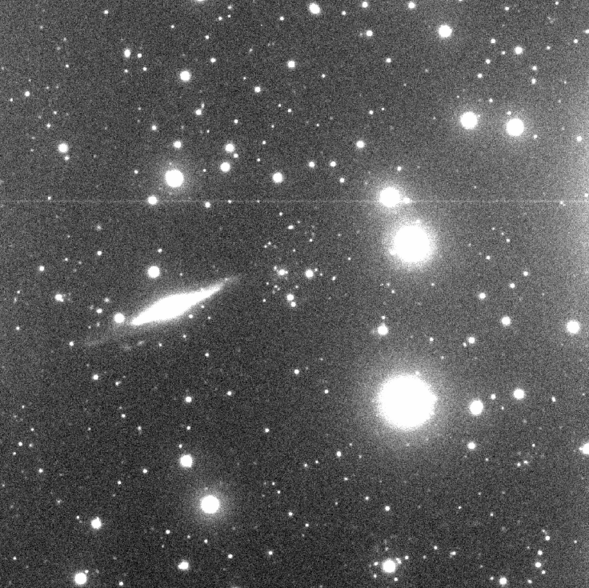} \\[\abovecaptionskip]
  \end{tabular}
  \hfill
  % \vspace{10pt}
  \begin{tabular}{@{}c@{}}
    \includegraphics[width=.24\linewidth]{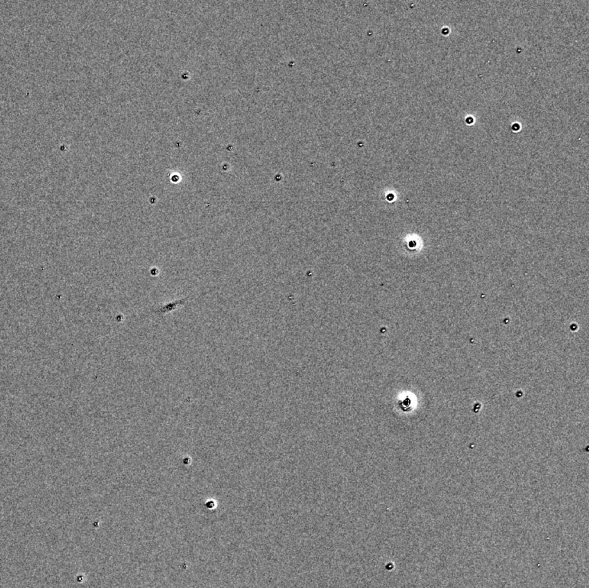} \\[\abovecaptionskip]
  \end{tabular}
    \hfill
  \begin{tabular}{@{}c@{}}
    \includegraphics[width=.24\linewidth]{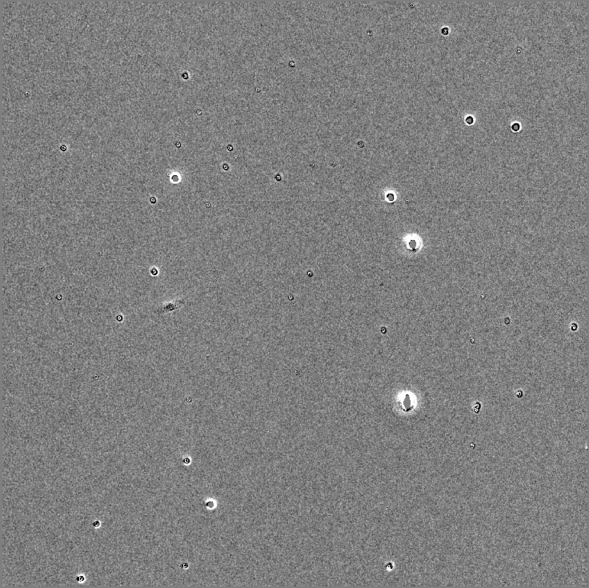} \\[\abovecaptionskip]
  \end{tabular} \\
    \centering
  \begin{tabular}{@{}c@{}}
    \includegraphics[width=.24\linewidth]{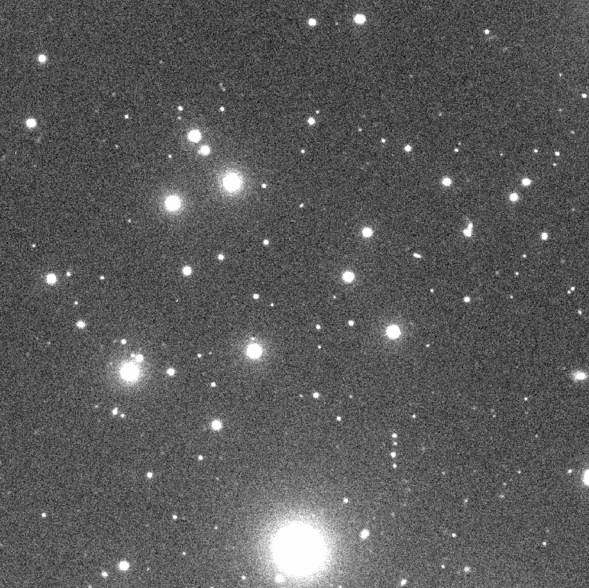} \\[\abovecaptionskip]
  \end{tabular}
  \hfill
  \begin{tabular}{@{}c@{}}
    \includegraphics[width=.24\linewidth]{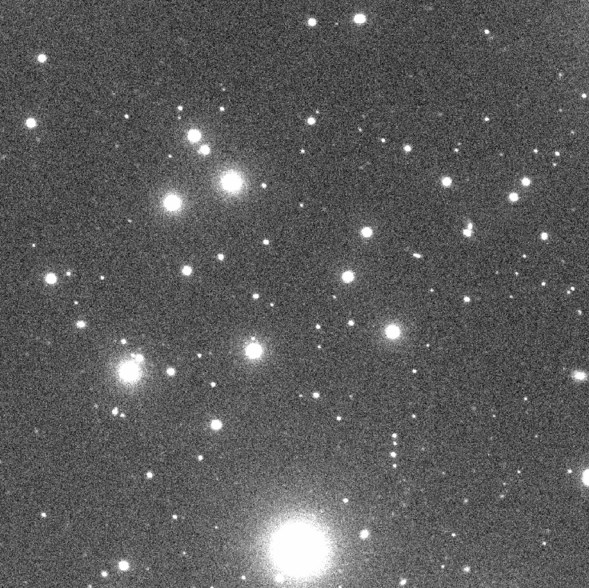} \\[\abovecaptionskip]
  \end{tabular}
  \hfill
  % \vspace{10pt}
  \begin{tabular}{@{}c@{}}
    \includegraphics[width=.24\linewidth]{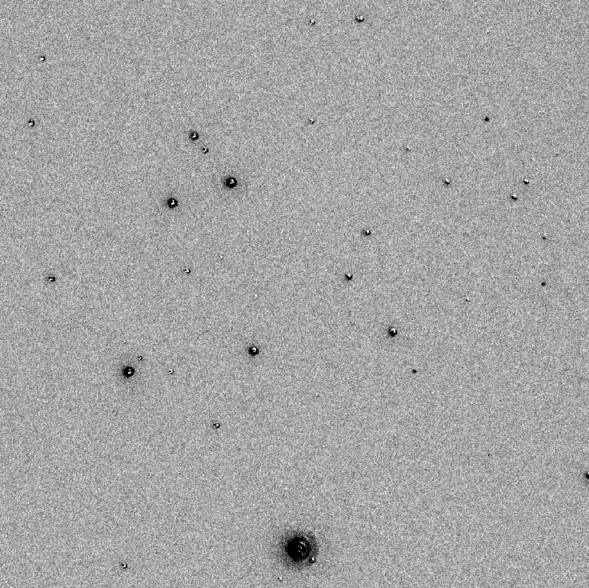} \\[\abovecaptionskip]
  \end{tabular}
    \hfill
  \begin{tabular}{@{}c@{}}
    \includegraphics[width=.24\linewidth]{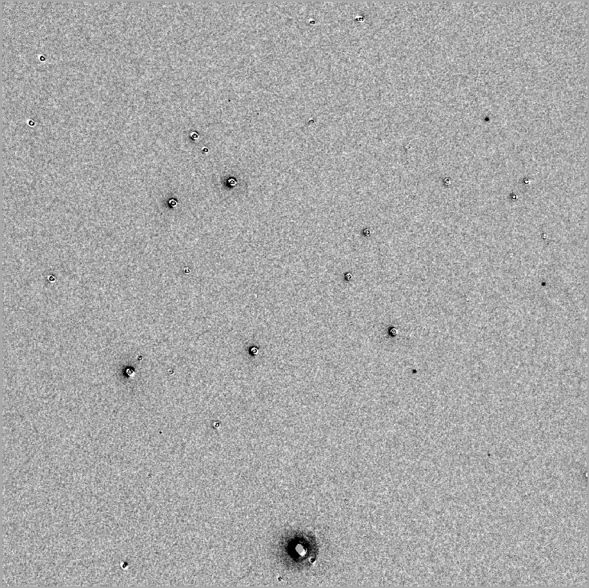} \\[\abovecaptionskip]
  \end{tabular} \\
  \centering
  \begin{tabular}{@{}c@{}}
    \includegraphics[width=.24\linewidth]{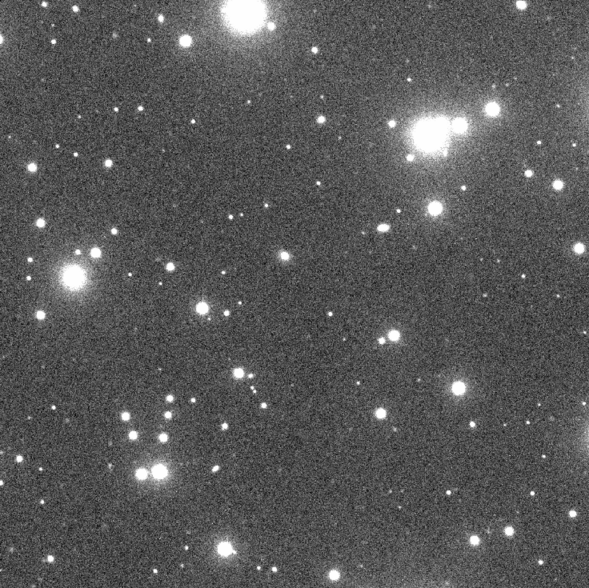} \\[\abovecaptionskip]
    \small (a) Science image
  \end{tabular}
  \hfill
  \begin{tabular}{@{}c@{}}
    \includegraphics[width=.24\linewidth]{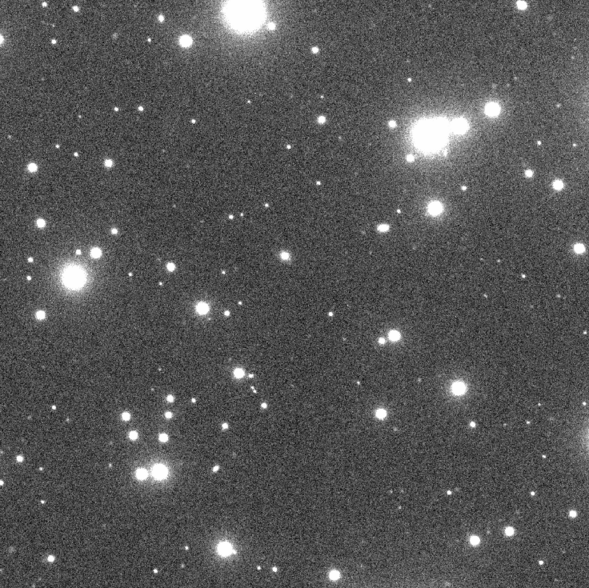} \\[\abovecaptionskip]
    \small (b) Reference image
  \end{tabular}
  \hfill
  % \vspace{10pt}
  \begin{tabular}{@{}c@{}}
    \includegraphics[width=.24\linewidth]{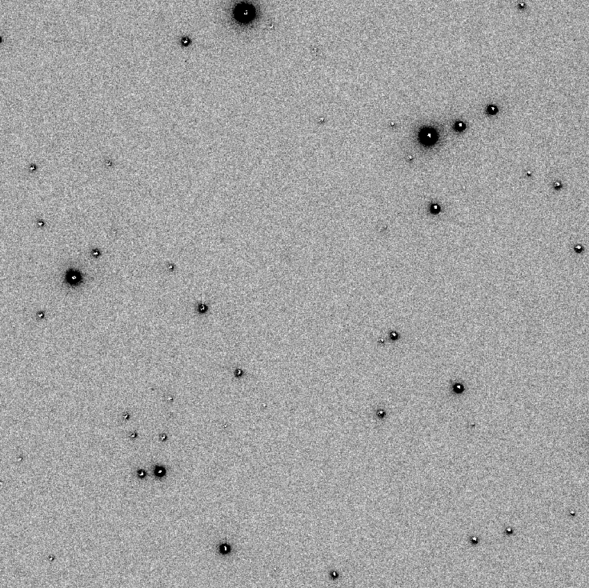} \\[\abovecaptionskip]
    \small (c) Subtracted image with \texttt{ILMTDiff}
  \end{tabular}
    \hfill
  \begin{tabular}{@{}c@{}}
    \includegraphics[width=.24\linewidth]{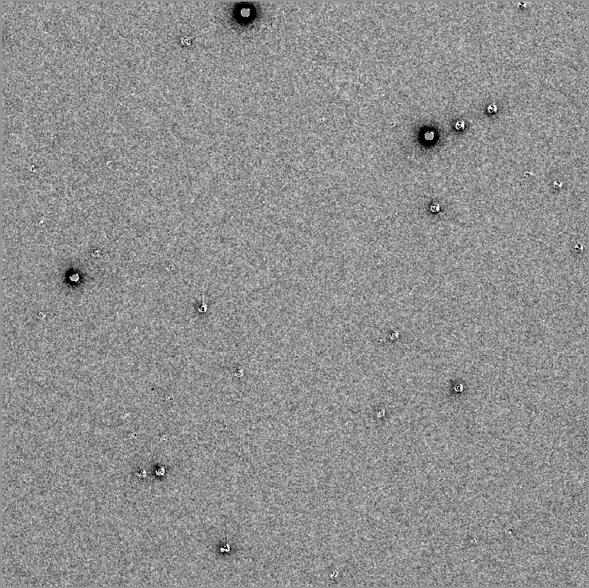} \\[\abovecaptionskip]
    \small (c) Subtracted image with \texttt{HOTPANTS}
  \end{tabular}
  \caption{Image subtraction was performed on 2K$\times$2K pixel ($\sim11' \times 11'$) cutouts of (a) science and (b) reference images of the ILMT using (c) \texttt{ILMTDiff} and (d) \texttt{HOTPANTS}. It can be inferred from the illustration that the quality of image subtraction with the custom \texttt{ILMTDiff} algorithm and with the \texttt{HOTPANTS} software are comparable. The presence of artefacts is noticeable in subtraction performed using either of the methods.}
\label{fig:comparative_subtraction}
\end{figure*}

\subsection{\texttt{ILMTDiff}}
The two aspects that the \texttt{ILMTDiff} image subtraction module was examined for were execution speed and quality of subtraction. The execution time for the \texttt{ILMTDiff} was optimised by fine-tuning the parameters to adapt to the ILMT images. The average processing time for 4096$\times$4096 pixel cutouts of ILMT image was $\sim$16--20 seconds. For comparison, the High Order Transform of Psf ANd Template Subtraction code (\texttt{HOTPANTS}) image subtraction software \citep{2015ascl.soft04004B} took similar time scales when tested on those images. It was also observed that the quality of subtracted images rendered by \texttt{ILMTDiff} was comparable to that by the \texttt{HOTPANTS} software (Figure~\ref{fig:comparative_subtraction}) for most pairs of images. Comparable performances were obtained for \texttt{Optimal Image Subtraction} \citep[\texttt{OIS};][]{OIS} and \texttt{Properimage} \citep{Properimage} which are two publically available \texttt{Python} based image subtraction codes. However, it should be noted that the performance of \texttt{ILMTDiff} depends on the choice of module parameters, which were determined after sufficient experimentation. In addition to the \texttt{ILMTDiff}, these other available codes can be integrated with the pipeline in future with an option to select among them. 

\subsection{\texttt{TransiSearch}}
\label{results_transisearch}

The evaluation of the \texttt{TransiSearch} module involved obtaining performance metrics for detecting transient/variable sources in the ILMT images. The adaptive detection technique described in Section~\ref{sec:adaptive_detection} was implemented to mitigate the false positive detection. For a quantitative evaluation of the effectiveness of this technique, the pipeline was implemented on 30 \textit{i}$'$ band images acquired on 28\textsuperscript{th} October, 29\textsuperscript{th} October and 30\textsuperscript{th} October 2022, both with and without implementing the adaptive detection technique.  

The version of the pipeline without implementing the technique produced 488 detections, of which 318 were true positives and 170 were false positives, resulting in a precision of 0.65. In contrast, the adaptive detection method yielded 320 detections, with 290 true positives and 30 false positives, yielding a precision of 0.91. It should be emphasised that the precision in this context differs from one calculated using an artificially curated and balanced \textit{test} dataset. In transient survey images like the ILMT, the ratio of the real to bogus samples is highly imbalanced as the number of bogus samples outnumbers the real transients by a significant margin.

\begin{figure*}
  \centering
  \begin{tabular}{@{}c@{}}
    \includegraphics[width=.49\linewidth]{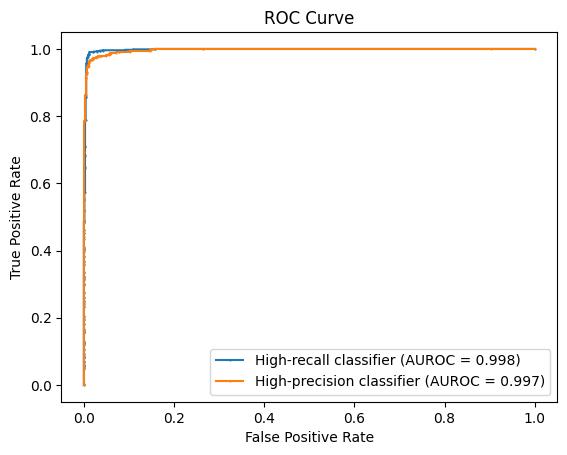} \\[\abovecaptionskip]
    % \small (a) ROC curves
  \end{tabular}
  \hfill
  \begin{tabular}{@{}c@{}}
    \includegraphics[width=.49\linewidth]{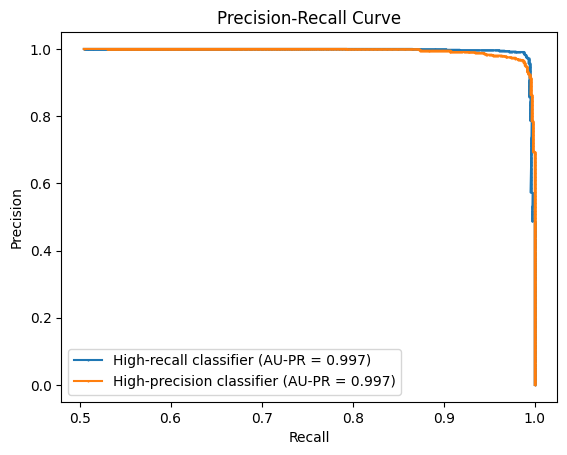} \\[\abovecaptionskip]
    % \small (b) PR curve
  \end{tabular}
  \caption{The ROC curves and PR curves for high-precision and high-recall CNN classifiers.}
\label{fig:ROC_PR}
\end{figure*}

\begin{figure}
  \centering
  \begin{tabular}{c@{\hspace{0.1cm}}c@{\hspace{0.1cm}}c@{\hspace{0.1cm}}c}
    \textbf{Science} & \textbf{Reference} & \textbf{Difference} & \textbf{Score} \\
    \includegraphics[width=0.25\linewidth]{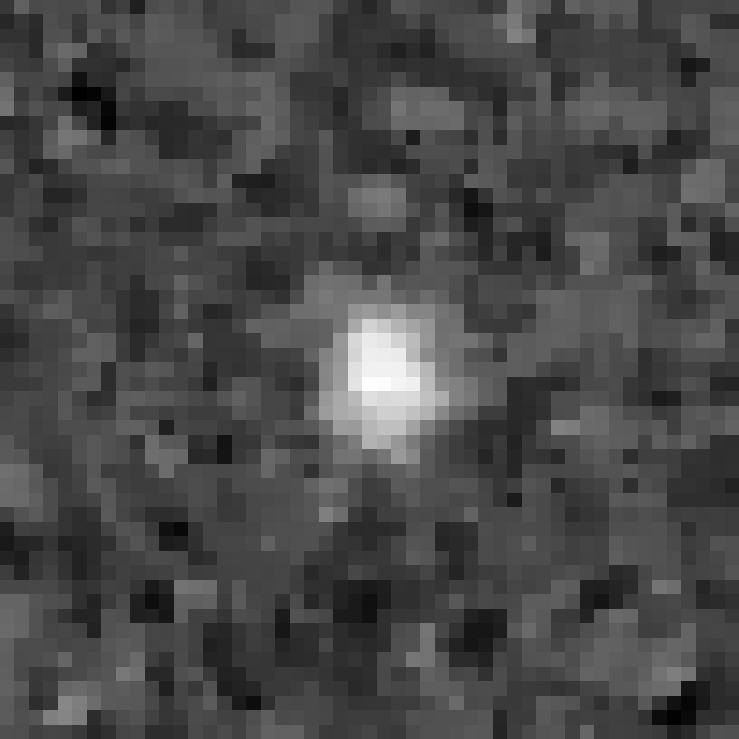} &
    \includegraphics[width=0.25\linewidth]{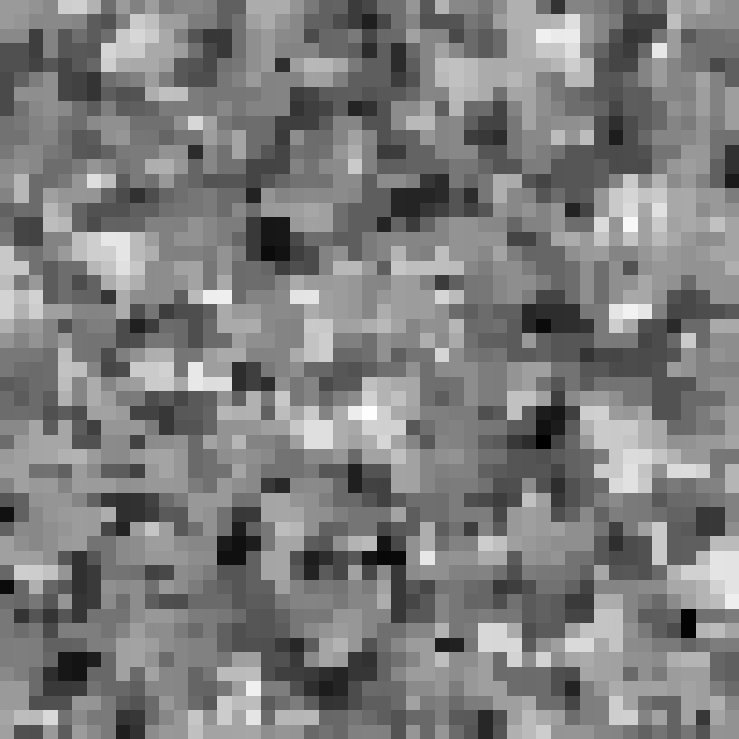} &
    \includegraphics[width=0.25\linewidth]{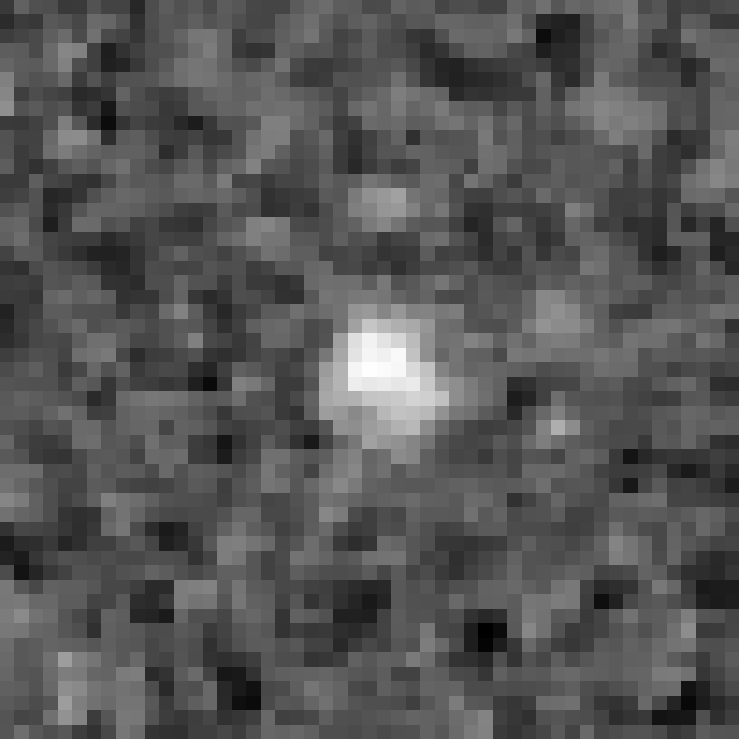} &
    \raisebox{0.8cm}{\textbf{0.9959813}} \\ 
    \includegraphics[width=0.25\linewidth]{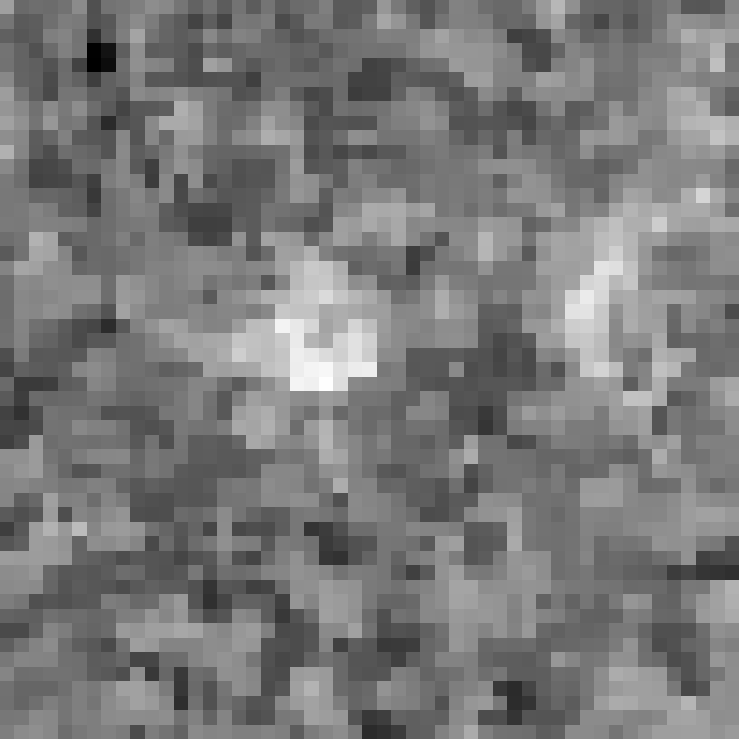} &
    \includegraphics[width=0.25\linewidth]{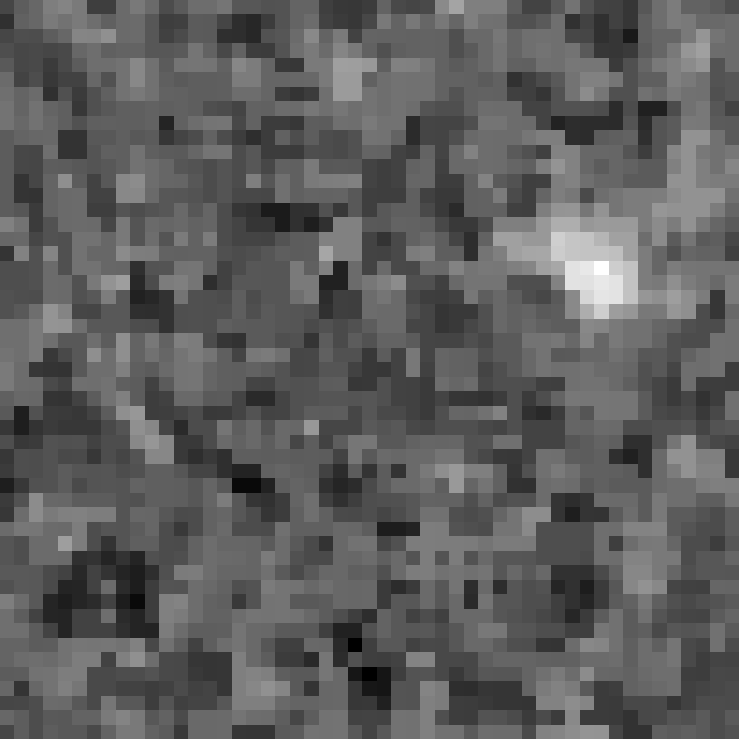} &
    \includegraphics[width=0.25\linewidth]{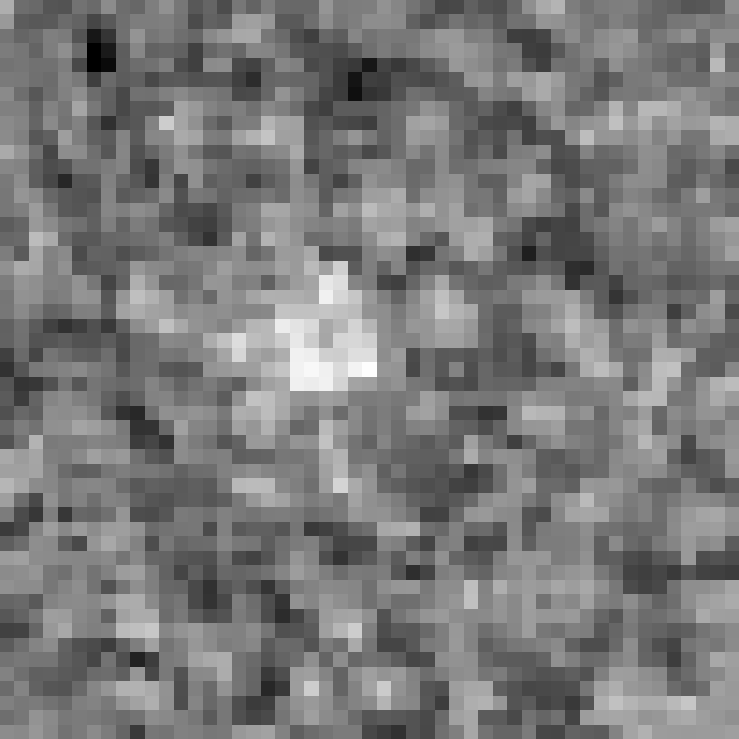} &
    \raisebox{0.8cm}{\textbf{0.7399178}} \\
    \includegraphics[width=0.25\linewidth]{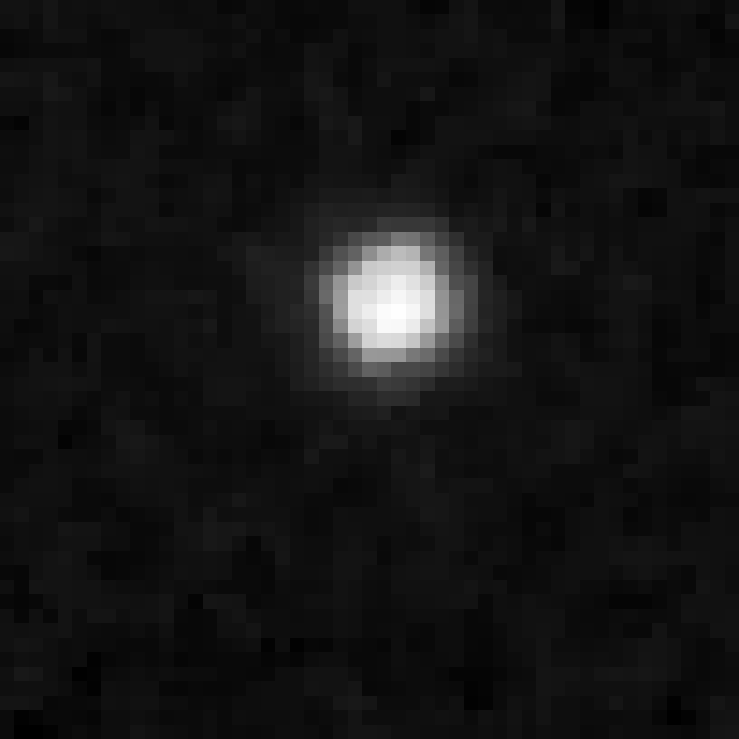} &
    \includegraphics[width=0.25\linewidth]{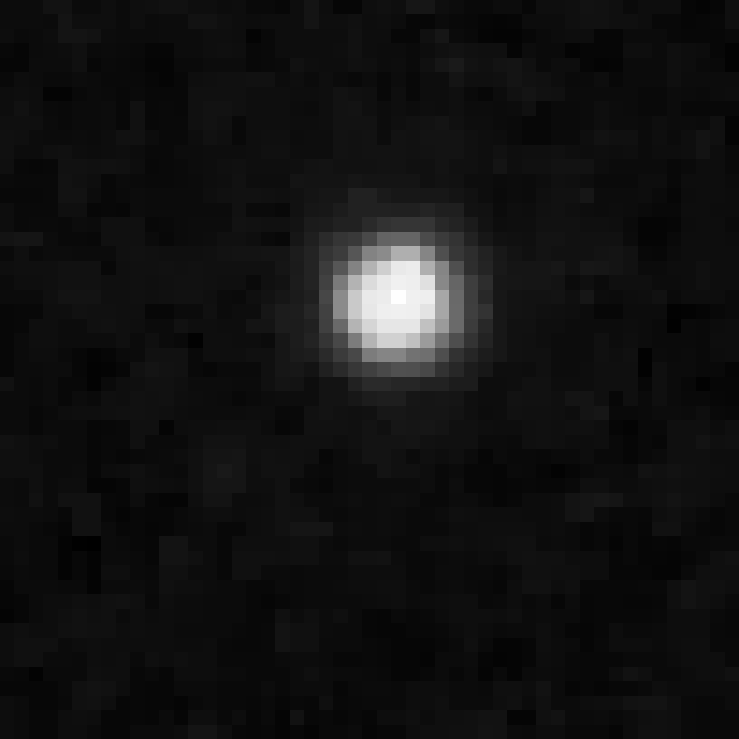} &
    \includegraphics[width=0.25\linewidth]{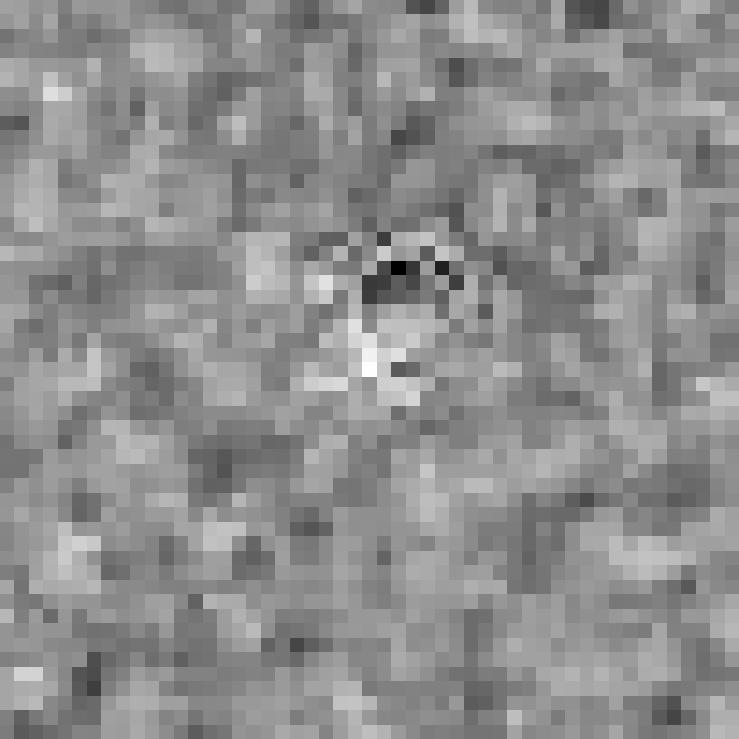} &
    \raisebox{0.8cm}{\textbf{6.3049274e-15}} \\
    \includegraphics[width=0.25\linewidth]{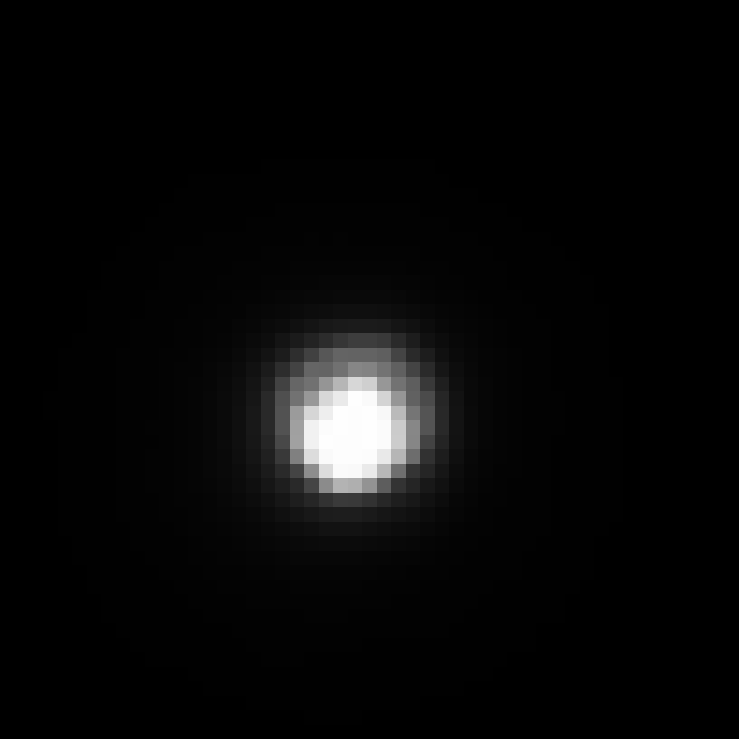} &
    \includegraphics[width=0.25\linewidth]{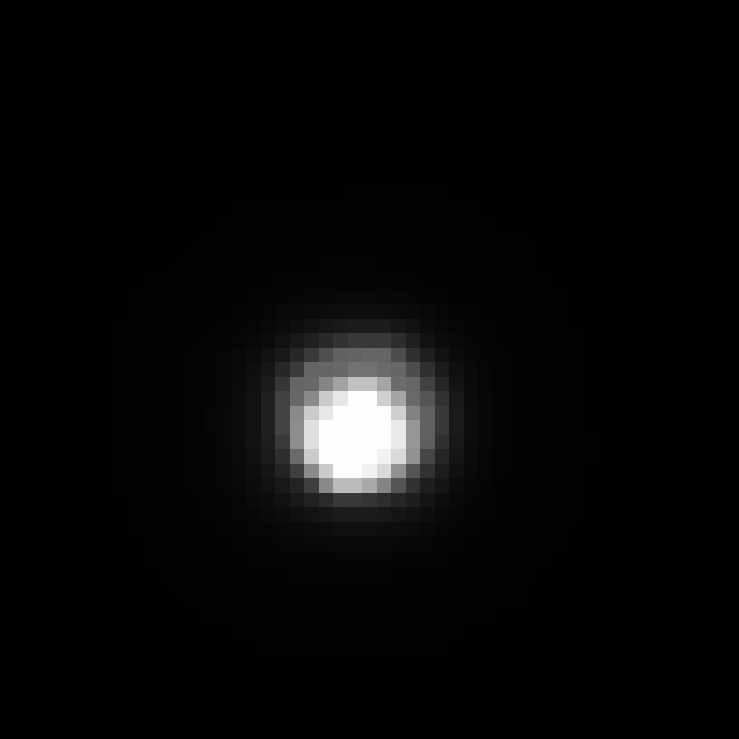} &
    \includegraphics[width=0.25\linewidth]{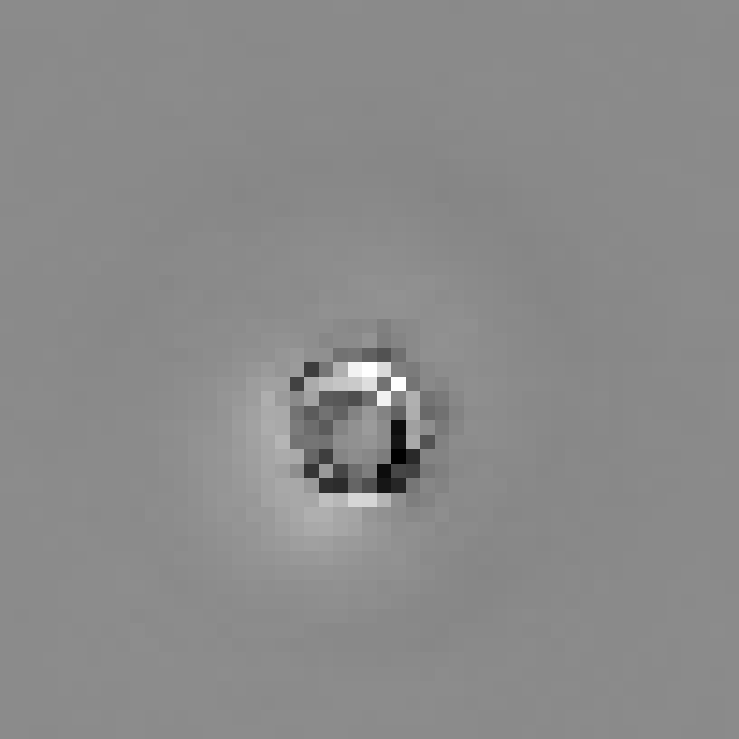} &
    \raisebox{0.8cm}{\textbf{1.9874052e-35}} \\
    \includegraphics[width=0.25\linewidth]{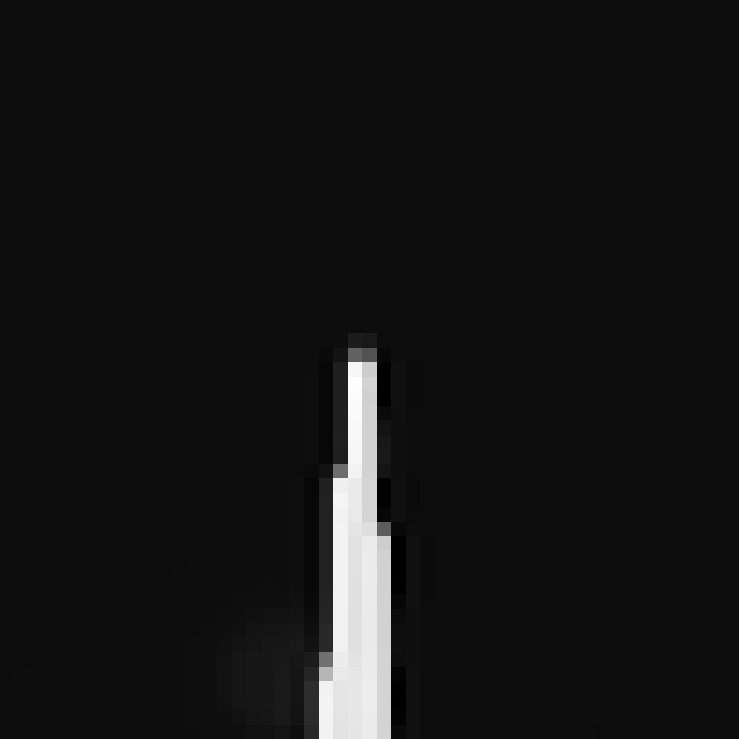} &
    \includegraphics[width=0.25\linewidth]{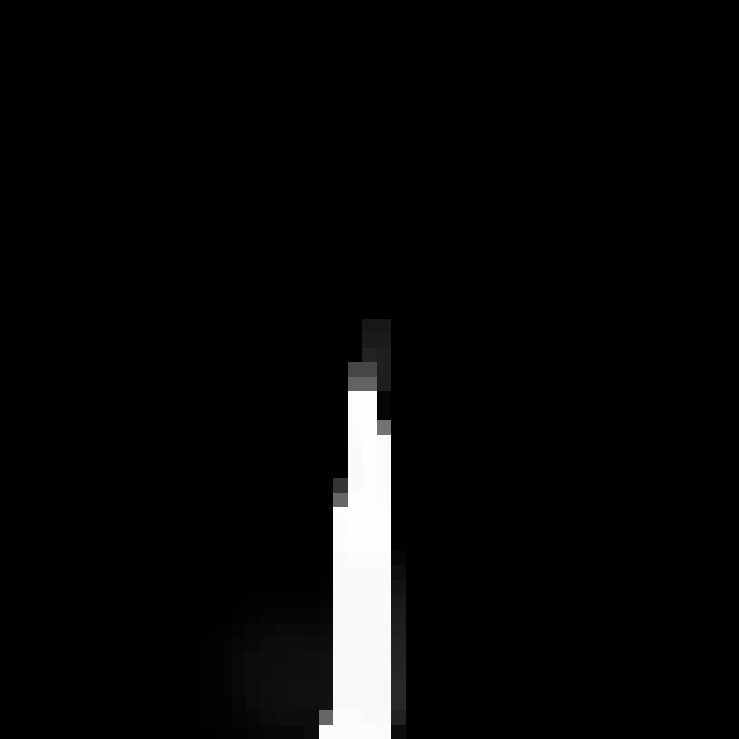} &
    \includegraphics[width=0.25\linewidth]{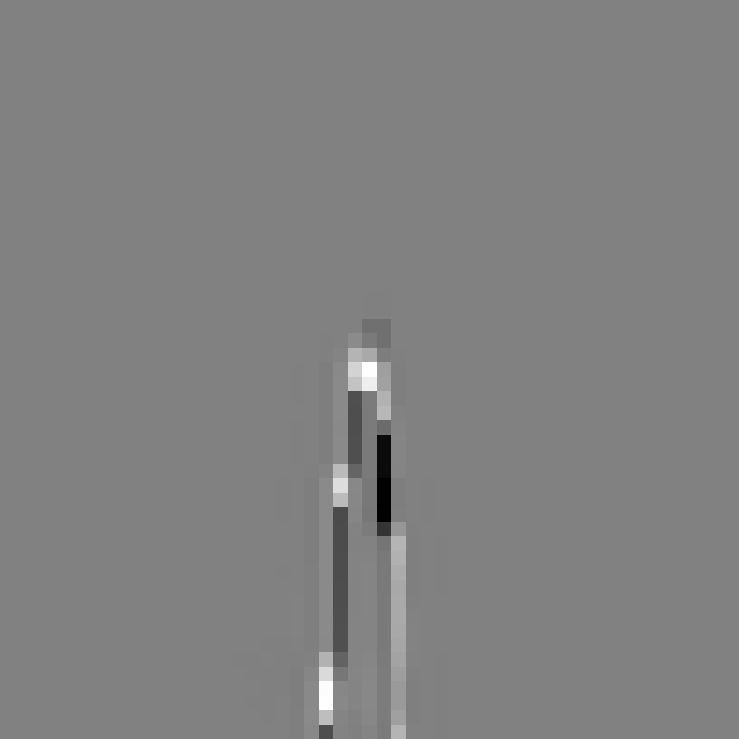} &
    \raisebox{0.8cm}{\textbf{4.3073076e-05}} \\
    \includegraphics[width=0.25\linewidth]{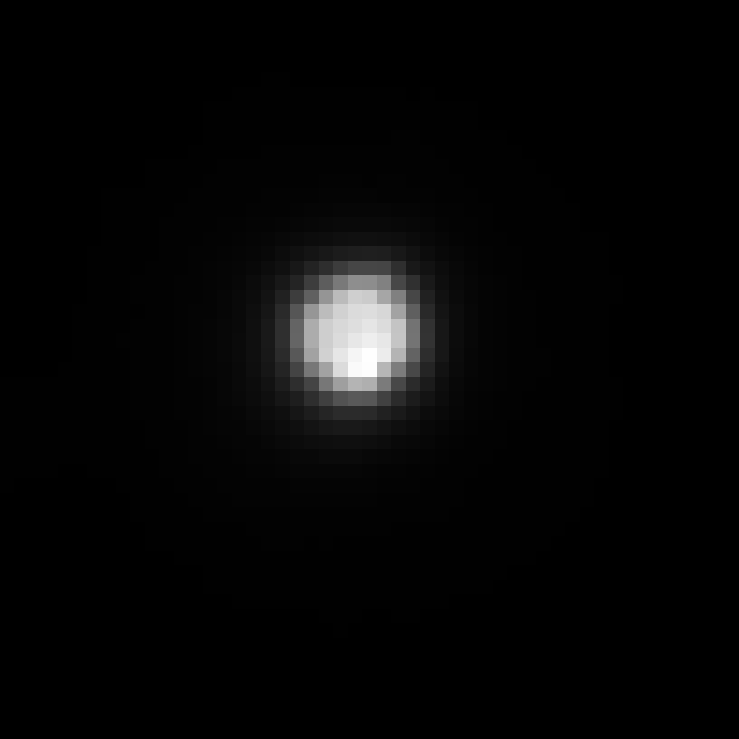} &
    \includegraphics[width=0.25\linewidth]{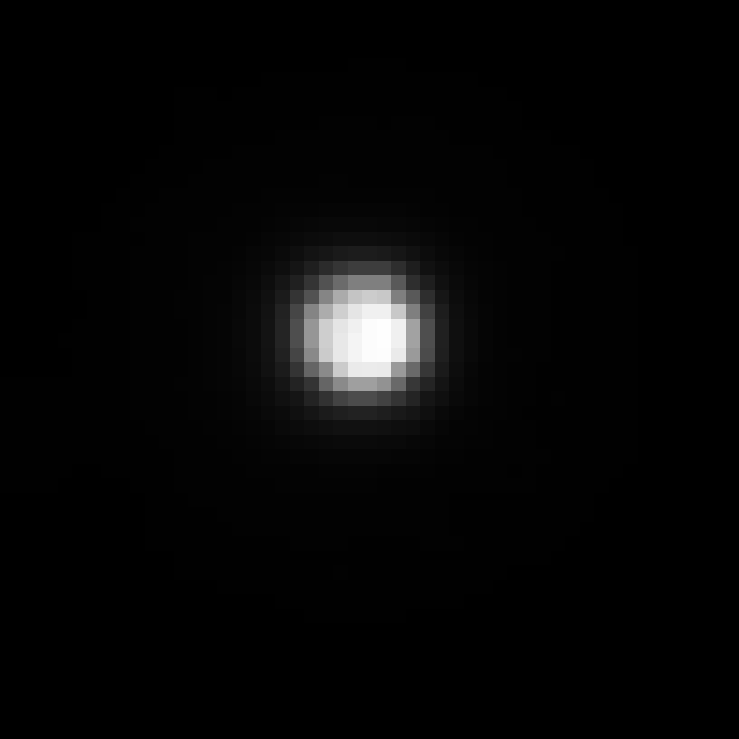} &
    \includegraphics[width=0.25\linewidth]{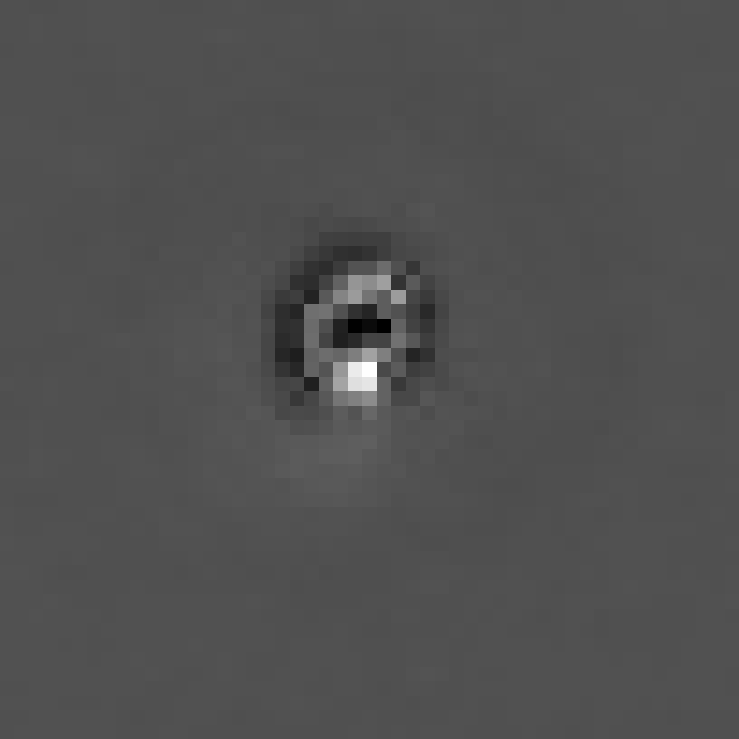} &
    \raisebox{0.8cm}{\textbf{7.924038e-06}} \\
  \end{tabular}
  \caption{51$\times$51 pixel ($\sim16''.5 \times 16''.5$) cutouts of sources in subtracted images along with their science and reference image cutouts and the corresponding \textit{real/bogus} classifier scores.}
  \label{tab:rbtable}
\end{figure}

The two CNN models employed for the adaptive detection technique were separately evaluated. The validation accuracy of high-precision and high-recall CNN classifiers were 98.96$\%$ and 97.71$\%$, respectively. The receiver operating characteristic (ROC) curve and the precision-recall (PR) curve for the two CNN models are shown in Figure~\ref{fig:ROC_PR}. The area under the ROC curve (AUROC) score for the high-recall classifier was determined to be 0.998, while that for the high-precision classifier was 0.997. The area under the PR curve (AU-PR) score for the high-recall classifier was 0.997, and that for the high-precision classifier was also 0.997. Confusion matrices for the high-recall and high-precision classifiers are given in Tables~\ref{tab:hr_confusion_matrix} and \ref{tab:hp_confusion_matrix}, respectively. The test accuracies were determined to be 98.07$\%$ and 93.97$\%$, respectively. The confusion matrices, the ROC, and the PR curves were generated using a separate test dataset with artificially added noise. Figure~\ref{tab:rbtable} lists some candidates with corresponding CNN scores. 

\begin{figure}
    \centering
    \includegraphics[width=0.47\textwidth]{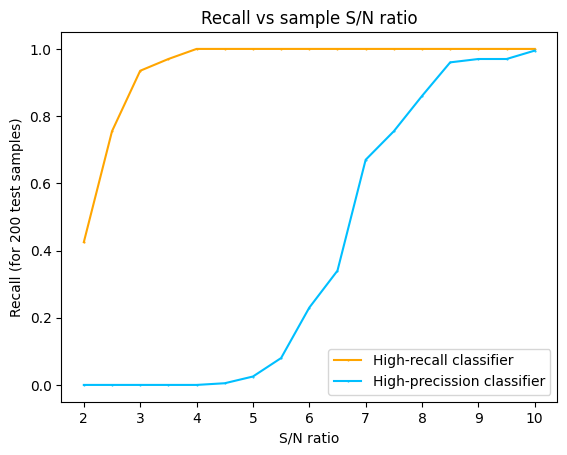}
    \caption{S/N ratio vs recall plot for high precision and high recall CNN-based real/bogus classifier.}
    \label{fig:recall-snr}
\end{figure}

The sensitivity of transient detection can be evaluated by determining recall values for `real' test candidates with varying S/N ratios. The test candidates were simulated with Gaussian profiles, by varying FWHMs and varying levels of artificial noise (to control the S/N ratio). All the candidates were then passed through the `real/bogus' classifier and threshold-based filtering. The mean recall for 200 candidates per S/N ratio was plotted and is shown in Figure~\ref{fig:recall-snr}. As expected, the high-recall classifier plateaued towards high-recall values for low S/N ratio samples, while the high-precision classifier showed a relatively steady increase.         

\begin{figure}
    \centering
    \includegraphics[width=0.47\textwidth]{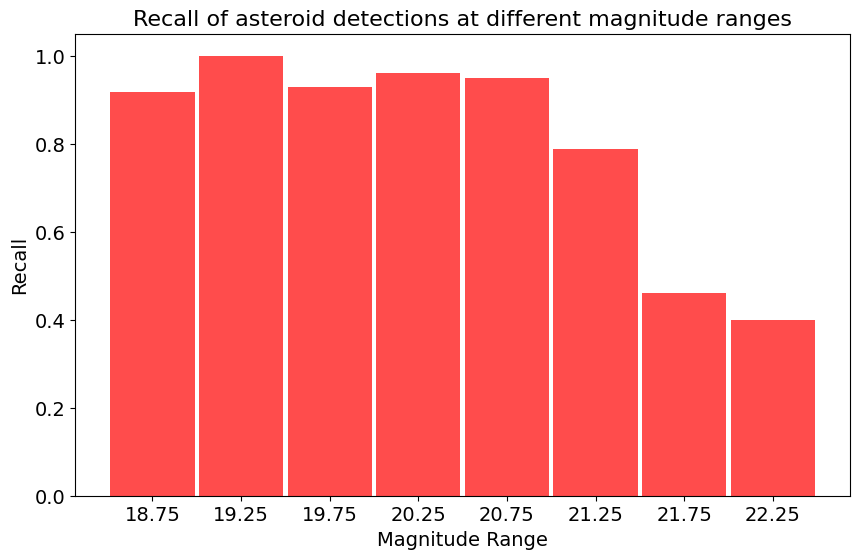}
    \caption{Recall for asteroid detection at different V-band magnitude ranges, with magnitudes taken from the Minor Planet Center (MPC).}
    \label{fig:recall-asteroid}
\end{figure}

\begin{figure}
    \centering
    \includegraphics[width=0.47\textwidth]{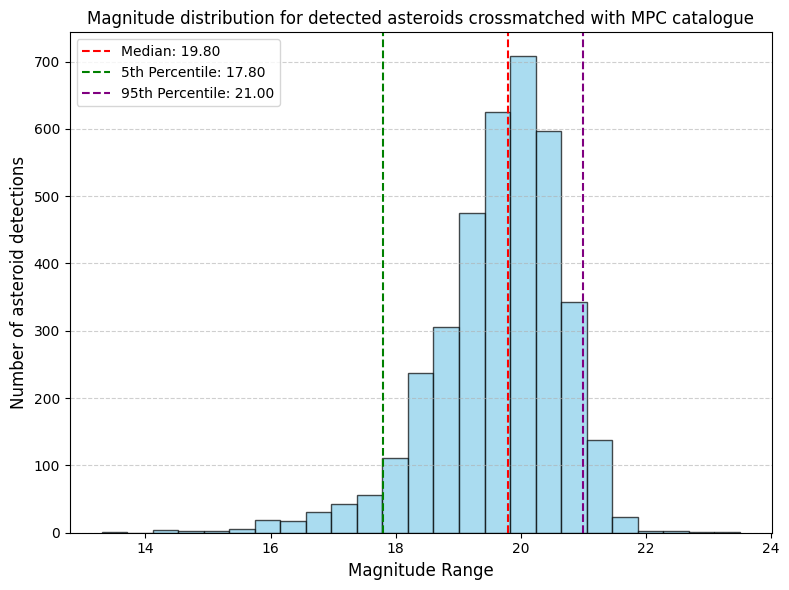}
    \caption{Distribution of MPC obtained V-band magnitudes of asteroids detected with the pipeline.}
    \label{fig:mag-asteroid}
\end{figure}

To evaluate the sensitivity of the overall pipeline on actual images, its detection completeness for visually inspected asteroids in 6 ILMT frames was determined. The V-band magnitudes of the detections were queried from the MPC catalogue of minor planets. A total of 293 MPC asteroids were confirmed visually. 159 of them were fainter than 21 magnitude. 231 of the 293 asteroids were successfully recovered by the pipeline. Figure~\ref{fig:recall-asteroid} illustrates the recall values for different V-band magnitude ranges. Furthermore, Figure~\ref{fig:mag-asteroid} presents the MPC obtained V-band magnitude distribution of the asteroids identified in real-time with the pipeline during its operation. From the distribution, the median V-band magnitude of the detected asteroids was determined to be 19.80, with the 5\textsuperscript{th} percentile magnitude at 17.80 and the 95\textsuperscript{th} percentile magnitude at 21.0.  

\begin{table}
\setlength{\extrarowheight}{5pt}
\centering
\caption{Confusion matrix for the high-recall classifier.}
\Large
\hspace{-50pt}
\begin{tabular}{cc|c|c|}
  & \multicolumn{1}{c}{} & \multicolumn{2}{c}{Predicted} \\
  & \multicolumn{1}{c}{} & \multicolumn{1}{c}{Bogus}  & \multicolumn{1}{c}{Real} \\\cline{3-4}
  \hspace{-5pt}
  \multirow{2}{*}{\rotatebox[origin=c]{90}{\hspace{-20pt} Actual}} & Bogus & 723 & 24 \\[10pt] \cline{3-4}
                          & Real & 5 & 756 \\[10pt] \cline{3-4}
\end{tabular}
\label{tab:hr_confusion_matrix}
\end{table}

\begin{table}
\setlength{\extrarowheight}{5pt}
\centering
\caption{Confusion matrix for the high-precision classifier.}
\Large
\hspace{-50pt}
\begin{tabular}{cc|c|c|}
  & \multicolumn{1}{c}{} & \multicolumn{2}{c}{Predicted} \\
  & \multicolumn{1}{c}{} & \multicolumn{1}{c}{Bogus}  & \multicolumn{1}{c}{Real} \\\cline{3-4}
  \hspace{-5pt}
  \multirow{2}{*}{\rotatebox[origin=c]{90}{\hspace{-20pt} Actual}} & Bogus & 744 & 3 \\[10pt] \cline{3-4}
                          & Real & 88 & 673 \\[10pt] \cline{3-4}
\end{tabular}
\label{tab:hp_confusion_matrix}
\end{table}

\subsection{\texttt{NovaNet}}

To evaluate the \texttt{NovaNet} module, classifications performed only on valid detections (as determined by visual inspection) by the \texttt{TransiSearch} module were considered. Table~\ref{tab:novanet} illustrates the predicted class distribution for valid detections with corresponding true positive class distribution at various common LST fields over 8 nights in October--November 2022. The test dataset was supplemented with a few artificially generated `SN-like' sources to account for the corresponding sample scarcity in the original dataset of detected sources. Such sources were generated by embedding point sources in a few galaxies in the ILMT frame. Table~\ref{tab:confusion_matrix} shows the confusion matrix for the classifications performed on the final test data. The classification accuracy was evaluated from the confusion matrix using the given formula and was determined to be 0.986.
\[
\text{Accuracy} = \frac{\text{$\Sigma$ diagonal elements}}{\text{$\Sigma$ all elements}}
\]

\subsection{Discovery of the new transients AT~2023yjc and 2024fxn}

\begin{figure*}
\setlength{\arrayrulewidth}{0.5mm}
    \centering
    \begin{tabular}{|c|c|}
        \hline
        \includegraphics[width=0.48\linewidth]{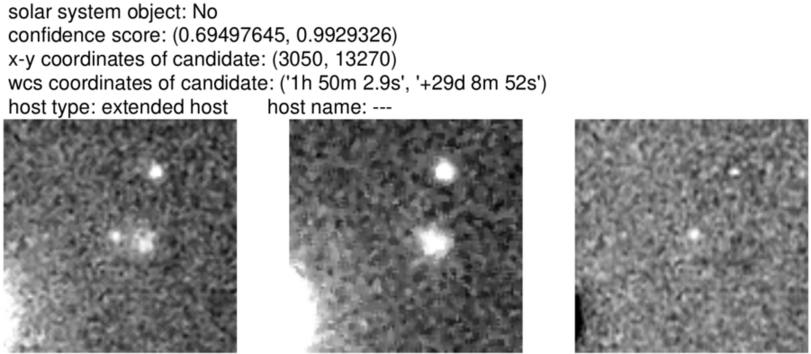} & \includegraphics[width=0.48\linewidth]{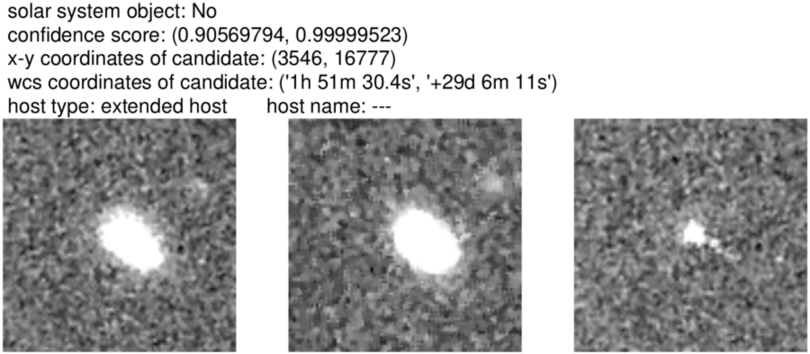}\\
        \hline
        \includegraphics[width=0.48\linewidth]{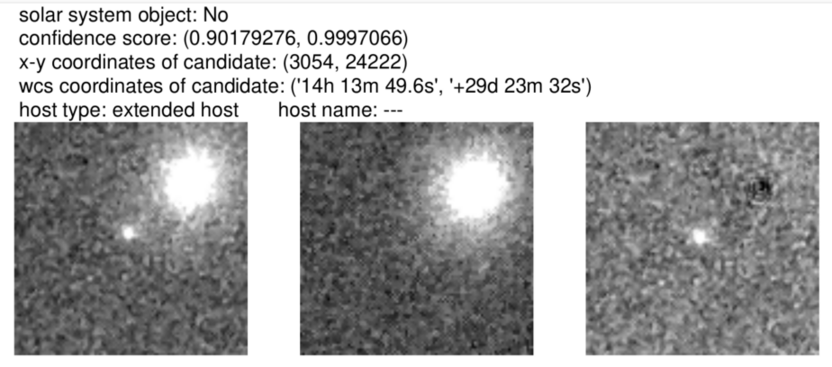} & \includegraphics[width=0.48\linewidth]{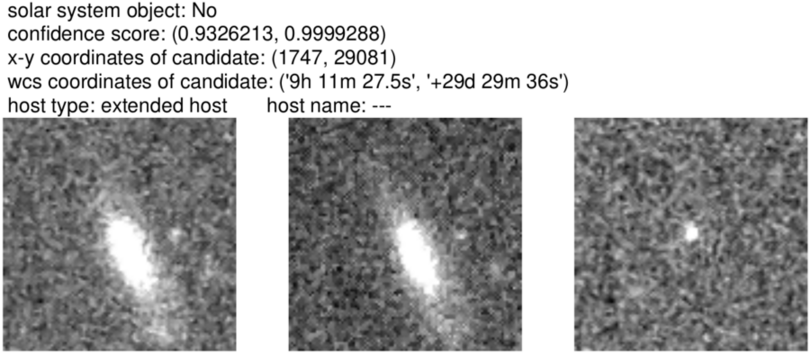}\\
        \hline
        \includegraphics[width=0.48\linewidth]{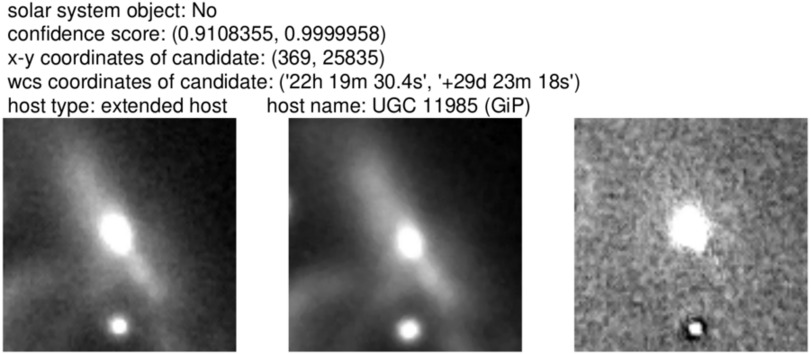} & \includegraphics[width=0.48\linewidth]{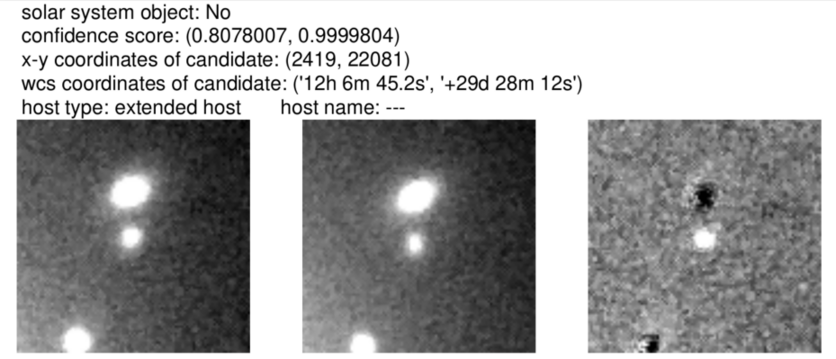}\\
        \hline
    \end{tabular}

    \caption{\textbf{Left column} -- top to bottom: \texttt{PyLMT} discovery of the new transients AT 2023yjc and 2024fxn on 13\textsuperscript{th} November 2023 and 5\textsuperscript{th} April 2024, respectively. The bottom panel illustrates the detection of the catalogued type-Ia SN~2023wuk. The images from left to right illustrate $\sim 33'' \times 33''$ cutout of the science image with the transient source, an older reference (or template) image and the subtracted (or difference) image resulting from the subtraction between the two. The results are in the adopted \texttt{PyLMT} alert format. \textbf{Right column} -- top to bottom: Detection of transients AT 2023xow on 11\textsuperscript{th} November 2023, SN~2024cjb on 14\textsuperscript{th} February 2024 and AT 2024ekk on 16\textsuperscript{th} March 2024.}
    \label{tab:detections_trans_1}
\end{figure*}

The third observation cycle of the ILMT commenced in November 2023. Using frames acquired in previous epochs as references, 12 `SN-like' transient events were detected using the pipeline in 761 analysed full-frame images from this cycle. The transients AT~2023yjc \citep{2023TNSTR3062....1P} and 2024fxn \citep{2024TNSTR.964....1P}, detected on 13\textsuperscript{th} November 2023 and 5\textsuperscript{th} April 2024, respectively, were reported as discoveries to the TNS. The names of the other detected transients include AT~2023vhj, 2023xow, 2024ccg, 2024ekk, 2024eab, and 2024fpx, alongside 4 confirmed SNe: SN~2023vcg, 2023wuk, 2023ngy, and 2024cjb. Six of the 12 detected transients are shown in Figure~\ref{tab:detections_trans_1}. 

\input{novanet}

\begin{table}
\setlength{\extrarowheight}{5pt}
\caption{Confusion matrix for the \texttt{NovaNet} candidate classifier.}
\begin{tabular}{cc|c|c|c|}
  & \multicolumn{1}{c}{} & \multicolumn{3}{c}{Predicted} \\
  & \multicolumn{1}{c}{} & \multicolumn{1}{c}{$ext.$}  & \multicolumn{1}{c}{$pt$}  & \multicolumn{1}{c}{$hostl.$} \\\cline{3-5}
            & $ext.$ & 14 & 0 & 1 \\[10pt] \cline{3-5}
Actual      & $pt$ & 5 & 185 & 5 \\[10pt]\cline{3-5}
            & $hostl.$ & 10 & 14 & 1284 \\[10pt]\cline{3-5}
\end{tabular}
\label{tab:confusion_matrix}
\end{table}

\section{Conclusions}
\label{sec:concl}

In this paper, we have discussed the \texttt{PyLMT} transient detection pipeline for the images acquired using the ILMT in the SDSS \textit{g}$'$, \textit{r}$'$, and \textit{i}$'$ spectral bands. The image subtraction is performed using an in-built module called \texttt{ILMTdiff} while transient detection and candidate classification are executed inside the \texttt{TransiSearch} and \texttt{NovaNet} modules, respectively. The pipeline is equipped with a catalogue cross-matching step to assist in the search for new transients by rejecting known asteroids and probable variable stars.

The quality of image subtraction achieved with the \texttt{ILMTDiff} module, based upon \citet{Bramich_2008} approach, is comparable with other existing software like the \texttt{HOTPANTS}. The precision value for real detections with the \texttt{TransiSearch} module, when implemented on ILMT frames, was determined as 0.91 and the test accuracy of the real/bogus classifiers ranged from 94\% to 98\%. Classification accuracy achieved with the \texttt{NovaNet} transient candidate classifier was 98.6$\%$. The pipeline has been successfully deployed on ILMT workstations and integrated with the data stream, leading to the detection of multiple transients, variable objects, and solar system bodies. The median execution time of the pipeline was determined to be approximately 29 minutes per ILMT frame. AT~2023yjc and 2024fxn were previously unreported transients discovered using the pipeline and subsequently reported to the TNS.

The ILMT science team regularly vets the candidates and interesting candidates are identified for follow-up. This potentially enables early-time identification of transients like SNe and their spectroscopic follow-up with other facilities. Such studies can be of great importance in enhancing our understanding of pre-explosion conditions of the SN progenitors, constraining the explosion mechanisms, and testing stellar evolution theories of massive stars.

\section*{Acknowledgements}
The authors thank the referee, Kendall Ackley, for providing constructive comments on the manuscript, which improved the clarity of the presentation. The 4-m International Liquid Mirror Telescope (ILMT) project results from a collaboration between the Institute of Astrophysics and Geophysics (the University of Li\`{e}ge, Belgium), the Universities of British Columbia, Laval, Montreal, Toronto, Victoria and York University, and Aryabhatta Research Institute of observational sciencES (ARIES, India). The authors thank Ankit Bisht, Hitesh Kumar, Himanshu Rawat, Khushal Singh, Nikhil Dharkiya and other observing staff for their assistance at the ILMT. The team acknowledges the contributions of AMOS (Advanced Mechanical and Optical Systems), CSL (Centre Spatial de Li\`{e}ge), Socabelec (Jemeppe-sur-Sambre) and ARIES's past and present scientific, engineering and administrative members in the realisation of the ILMT project. JS wishes to thank Service Public Wallonie, F.R.S.-FNRS (Belgium) and the University of Li\`{e}ge, Belgium for funding the construction of the ILMT. ARIES thanks the Department of Science and Technology (DST), Govt. of India, for the realisation of the project. PH acknowledges financial support from the Natural Sciences and Engineering Research Council of Canada, RGPIN-2019-04369. PH and JS thank ARIES for their hospitality during their visits to Devasthal. JS and KM acknowledge the assistance received from the Anusandhan National Research Foundation (ANRF, SERB- 762 VAJRA Faculty Scheme, India). KM, BK, BA and ND acknowledge the support from the BRICS grant DST/ICD/BRICS/Call-5/CoNMuTraMO/2023 (G) funded by the DST, India. BA acknowledges the Council of Scientific $\&$ Industrial Research (CSIR) fellowship award (09/948(0005)/2020-EMR-I) for this work. MD acknowledges the Innovation in Science Pursuit for Inspired Research (INSPIRE) fellowship award (DST/INSPIRE Fellowship/2020/IF200251) for this work. KP and KM would also like to thank Ashish Mahabal and Stefano Valenti for their valuable advice and insightful suggestions.

%%%%%%%%%%%%%%%%%%%%%%%%%%%%%%%%%%%%%%%%%%%%%%%%%%
\section*{Data Availability}

To perform this work, the authors extensively used the images acquired with the ILMT during its commissioning period. The raw and processed images with astrometric calibrations are routinely made available to the public domain and can be accessed using the URL \url{https://cloud.aries.res.in/index.php/s/xPER9Y3XuaCsTL9}. The publicly available survey images from the ZTF were used to perform initial testing of the pipeline scheme, including key algorithms like image subtraction. The images can be accessed through the IRSA platform using the URL \url{https://irsa.ipac.caltech.edu/applications/ztf/?__action=layout.showDropDown&}. Other relevant data can be made available upon request to the authors.     



\bibliographystyle{mnras}
\bibliography{Main_PyLMT} % if your bibtex file is called example.bib

%\begin{thebibliography}{99}
%\end{thebibliography}

%%%%%%%%%%%%%%%%%%%%%%%%%%%%%%%%%%%%%%%%%%%%%%%%%%

%%%%%%%%%%%%%%%%% APPENDICES %%%%%%%%%%%%%%%%%%%%%

\appendix

\section{The \texttt{ILMTDiff} image subtraction algorithm} \label{app:Image_subtraction}
The process of image subtraction involves a few basic steps namely image alignment, background subtraction, flux matching, and PSF matching. Each of these steps as implemented in the \texttt{ILMTDiff} module is discussed below in detail.

\begin{enumerate}
    \item \textbf{Science and Reference images:} The reference image has to be subtracted from the science image to obtain the difference image. Transient detection is performed on this difference image. The reference image required for image subtraction should be acquired on a night with relatively better seeing conditions. Co-added reference images are desired to produce subtracted images with good S/R ratio. Limited imaging data was available during the pipeline validation stage. Hence, the current work used non-coadded single reference images acquired in good seeing conditions. With the acquisition of more data in future, a database of co-added reference images will be constructed.

    \item \textbf{Image alignment:} The science and reference images may exhibit relative offsets due to imprecision in the timing of image acquisition. Additionally, due to the fixed-pointing nature of the telescope, the images are subject to Earth's precession and nutation effects, causing systematic shifts along RA and Dec. The relative offset becomes more pronounced with increasing duration between the acquisition of science and reference images. Thus, the images are first aligned with the \texttt{shift} module from the \texttt{scipy} library, based on the WCS information in the \texttt{FITS} headers. Inaccuracies in astrometric calibration further influence the alignment of the two images. To address this potential inaccuracy, the module has a provision to align images using \texttt{Astroalign} software \citep{beroiz2020astroalign}.     
    
    \item \textbf{Background subtraction:} The frames acquired with the ILMT exhibit uniform or spatially varying background illumination. This background is affected by factors such as moon phase, scattering of moonlight due to clouds, and even possible light leakage in the dome (now under control). These factors can cause a significant variation in background illumination across different ILMT frames. Therefore, the space-varying background is removed from the acquired images using the \texttt{MedianBackground} estimator available with \texttt{photutils} \citep{2016ascl.soft09011B}.

    \item \textbf{Flux matching:} Depending on factors like cloud cover, presence of haze, reflectivity, and transmissivity of the telescope components etc, the total instrumental flux integrated for the sources in the image might differ from one night to another. Image subtraction requires integrated source flux in science and reference images (acquired on different nights) to be matched. The kernel optimisation step in the \texttt{ILMTDiff} module itself scales the kernel with the appropriate flux scaling factor as a consequence of PSF matching. Nevertheless, provision has been made to calculate the scaling factor separately by evaluating the median of the aperture flux ratios of common sources in the science and reference images.

    \item \textbf{PSF matching:} PSF matching is the principle functional step of the image subtraction module. The images are acquired on different nights with possibly different seeing conditions. For successful image subtraction, it is crucial to match PSF between the sources in science and reference images. This is achieved by convolving the image with better seeing (smaller FWHM; ideally the reference image) with an appropriate convolution kernel. This kernel can be determined using least-squares optimisation, which in this context, involves minimising the sum of squares of pixel values for subtraction residuals in the difference images. The delta-basis formulation (refer to Section \ref{sec:delta-basis}) was used as the principle kernel model for subtraction.  
\end{enumerate}

\input{ILMTDiff_threshold}

\subsection{Delta-basis kernel optimisation for PSF matching} \label{sec:delta-basis}

The algorithm begins by selecting a set of common sources in the science and reference images. These sources are above a specified detection threshold and are uniformly distributed across the image. The sources are extracted in square cutouts of specified size, with a default value of 21$\times$21 pixels. This value can be adjusted with the \textit{stamp\_size} parameter. A 2-D Gaussian profile is fitted to these cutouts, allowing the determination of morphological parameters. A series of thresholds (refer to Table \ref{tab:ILMTDiff_threshold}) are then applied to these parameters to filter `good' sources for subsequent steps. The maximum number of sources used for kernel determination can be specified using the \textit{n\_stamps} parameter in \texttt{ILMTDiff}. A larger value for \textit{n\_stamps} parameter causes the algorithm to consume more time. Values ranging from 5--30 have been used for this parameter, yielding acceptable subtraction quality.  

The source cutouts in the reference image are convolved with a kernel with randomly initialised pixel values and are subtracted from the cutout of the corresponding source in the science image. The kernel size can be specified by the user (default value is set at 7$\times$7 pixels). The pixel values of the kernel are iteratively modified using a numerical optimisation method \citep{1978LNM...630..105M, Virtanen_2020} to get the optimal kernel. The sum of residuals of the subtraction between science image cutouts and convolved reference image cutouts is the objective function to be minimised. Mathematically, the residuals can be expressed using equations \ref{eq:stamp_residual} and \ref{eq:total_residual}.

\begin{equation} \label{eq:stamp_residual}
    stamp\_residual = \sum_{i,j}\left( \frac{S_{ij} - \sum_{\substack{k,l}} R_{(i+k)(j+l)} K_{kl}}{\sigma_{ij}} \right)^{2}
\end{equation}

\begin{equation} \label{eq:total_residual}
        total\_residual = \sum_{\substack{n=1}}^{\textit{n\_stamps}}stamp\_residual
\end{equation}

where $S_{ij}$, $R_{ij}$ are the pixel values for individual science and reference source cutouts, $K_{kl}$ represent the pixel values for the convolution kernel (in kernel coordinates) and $\sigma_{ij}$ is the total of background noise and Poisson noise contribution from science and convolved reference images. \textit{stamp\_residual} is the subtraction residual for an individual pair of science and reference source cutouts (or stamps) while \textit{total\_residual} is the sum of all \textit{stamp\_residuals}, which has to be minimised. Kernel optimisation automatically adjusts the kernel to account for the flux factor. In an optimised kernel, the sum of pixel values equals the flux scaling factor.

\begin{figure*}
    \centering
    \includegraphics[width=0.85\textwidth]{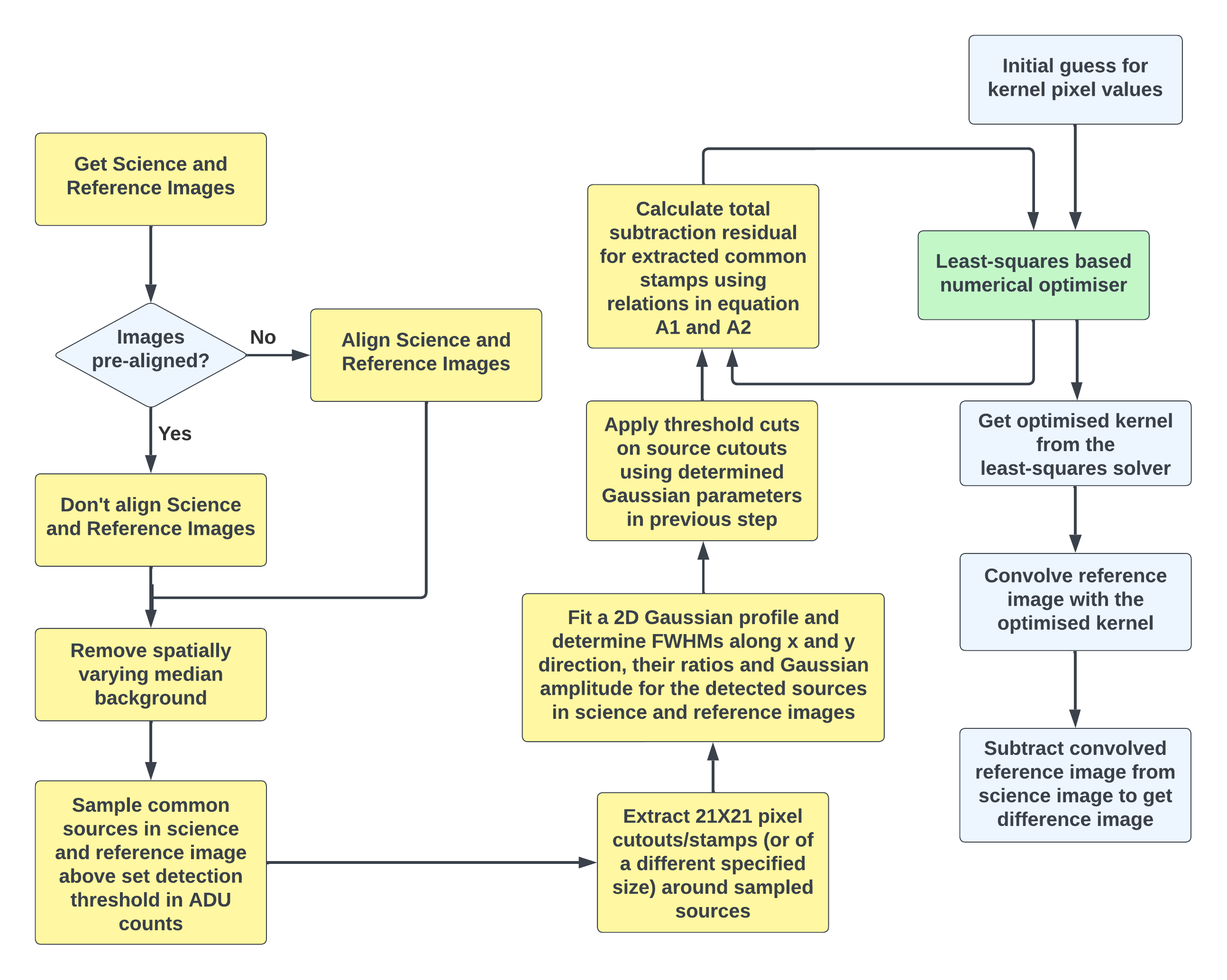}
    \caption{Flowchart of the \texttt{ILMTDiff} image subtraction module.}
    \label{fig:ILMTDiff_flowchart}
\end{figure*}

Upon determination of the optimal kernel, the reference image is convolved with this and is subtracted from the science image to get the subtracted/difference image. Figure \ref{fig:ILMTDiff_flowchart} illustrates the flowchart for the \texttt{ILMTDiff} image subtraction module.

\section{Training of the real/bogus classifier} \label{app:Real/Bogus}

\subsection{Dataset preparation}
It is discussed in Section \ref{sec:adaptive_detection} that high-precision and high-recall CNN classifiers are alternatively used to reduce false positive detections effectively. The class distribution of the dataset was kept different to construct the two different classifiers. The number of `real' and `bogus' samples was 2705 and 2527, respectively for the high-recall classifier. However, the same was 1000 and 2527, respectively for the high-precision classifier. A relatively greater number of `bogus' samples in the training dataset biased the high-precision classifier to be more sensitive in identifying bogus sources accurately and therefore have low overall false positive detections.

The ILMT data from the first commissioning cycle in October-November 2022 was used to construct the dataset. The samples are square cutouts of size 31$\times$31 pixels, corresponding to nearly 10$\times$10 arcsecond\textsuperscript{2} of the sky. The astrometric seeing of the ILMT can range from 1$''$.5 to 3$''$.5. Also, there are occasional appearances of comatic flaring and astrometric misalignment of images (up to 1$''$). The cutout size is sufficiently large to encapsulate the size and nature of the sources. Simultaneously, it is sufficiently small to constrain individual sources in the difference source cutouts. The real sources were obtained by extracting cutouts of `PSF-like' point sources from acquired ILMT frames. Bogus samples were extracted from subtracted ILMT frames. Data augmentation by orthogonal rotations was utilised to enhance the model's generalisability and the extensiveness of the dataset. Special care was taken to ensure diversity in the visual appearances of the dataset samples, enabling the training of robust networks.   

\subsection{Data pre-processing and training}
The samples in the dataset had a wide distribution in scales of pixel values depending on source brightness. The models trained with such a dataset can be unstable and perform poorly on real data. Therefore, it was decided to standardise dataset samples using Equation \ref{eq:standardization}. 
\begin{equation} \label{eq:standardization}
    Z = \frac{{X - \mu}}{{\sigma}}
\end{equation}
Where:
\begin{align*}
    Z & \text{ is the standardised sample,} \\
    X & \text{ is the raw sample,} \\
    \mu & \text{ is the mean of the sample pixels,} \\
    \sigma & \text{ is the standard deviation of the sample pixel values.}
\end{align*}

Artificial noise was added to some of the dataset samples to enhance the detection sensitivity of the trained model. The dataset was split into training and validation sets in an 80:20 ratio. 

The same CNN architecture was used for the high-precision and high-recall classifiers. The architecture consists of a feature extraction region with 3 convolutional layers, each coupled with a 2$\times$2 maxpooling. The model uses a combination of 2$\times$2 and 3$\times$3 pixel size filters. The fully connected region has 2 hidden layers. The activation function used in each layer is \texttt{ReLU} \citep{inproceedings_nair} except for the output layer where \texttt{sigmoid} activation function is used. The complete CNN architecture is shown in Figure \ref{fig:CNN_archi_63}.

For training, the loss function used was \texttt{binary-crossentropy} and the optimiser was \texttt{adam} \citep{adam}. Regularisation techniques were also used to prevent overfitting. Two techniques used were \texttt{dropout} regularisation \citep{JMLR:v15:srivastava14a} and \texttt{L2} regularisation. Dropout rates of 0.5 in the feature extraction region and 0.7 in the fully connected region appeared to work well. The regularisation hyperparameter for the L2 regulariser was set to 0.02.

\begin{figure*}
  \centering
  \begin{tabular}{@{}c@{}}
    \includegraphics[width=.49\linewidth]{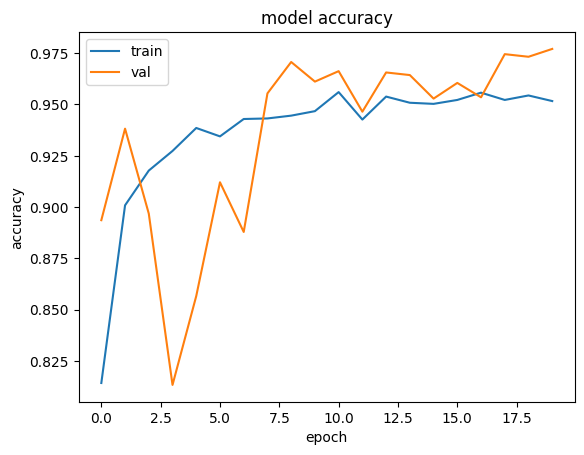} \\[\abovecaptionskip]
    % \small (a) accuracy curves
  \end{tabular}
  \hfill
  \begin{tabular}{@{}c@{}}
    \includegraphics[width=.49\linewidth]{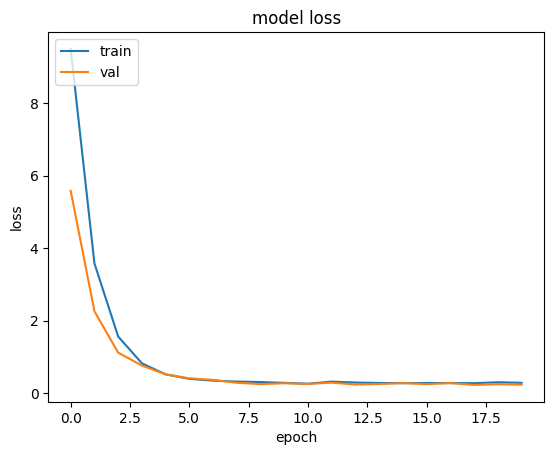} \\[\abovecaptionskip]
    % \small (b) loss curve
  \end{tabular}
  \caption{The accuracy and loss curves for high-recall CNN classifier.}
\label{fig:acc_loss_63}
\end{figure*}

\begin{figure*}
  \centering
  \begin{tabular}{@{}c@{}}
    \includegraphics[width=.49\linewidth]{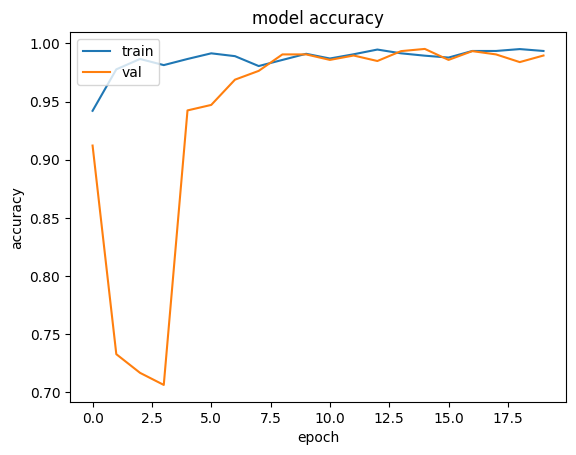} \\[\abovecaptionskip]
    % \small (a) accuracy curves
  \end{tabular}
  \hfill
  \begin{tabular}{@{}c@{}}
    \includegraphics[width=.49\linewidth]{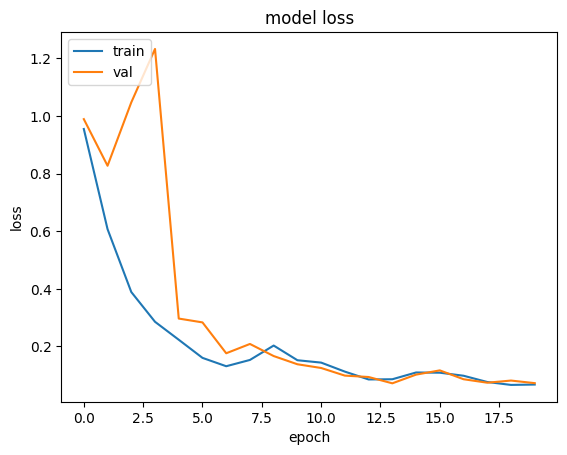} \\[\abovecaptionskip]
    % \small (b) loss curve
  \end{tabular}
  \caption{The accuracy and loss curves for high-precision CNN classifier.}
\label{fig:acc_loss_71}
\end{figure*}

Additional data augmentation was performed by applying \textit{rotation\_range}, \textit{horizontal\_flip}, \textit{height\_shift\_range} and \textit{width\_shift\_range} using \texttt{Keras} \texttt{ImageDataGenerator} module \citep{chollet2015keras}. Then CNN architecture was written and trained as a \textit{sequential} model using Google's \texttt{TensorFlow} \citep{abadi2016tensorflow} library. GPU mode available with Google Colab\textsuperscript{\tiny\textregistered} was used for training. After multiple instances of training, it was observed that the validation accuracy plateaued around 98$\%$ in about 20 epochs with a default batch size of 32. The accuracy and loss curves for the high-recall classifier are shown in Figure \ref{fig:acc_loss_63} and the same for the high-precision classifier are shown in Figure \ref{fig:acc_loss_71}.

\section{Training of the transient candidate classifier} \label{app:transient_classifier}
\subsection{Dataset preparation}
The dataset for the \texttt{NovaNet} transient candidate classifier was curated using the ILMT science frames acquired in the October-November cycle of 2022. The dataset was created by extracting cutouts from ILMT frames representing three different \textit{host} situations (refer to Section \ref{sec:class}): (i) `extended-host' (ii) `point-host' and (iii) `hostless'. 

The `extended-host' situation is characterised by a galaxy present in the reference image at the position of detection. `Point-host' is characterised by a point-source and `hostless' by the absence of any source. So to prepare the dataset, appropriate cutouts were cropped out from the ILMT images. For the `extended-host' samples, it was ensured that the displacement of the galaxies relative to the cutout centres mimicked the spatial occurrences of transients relative to its host galaxy.   

A total of 407 samples for each category were cropped out. The samples were augmented with 90\textdegree, 180\textdegree, and 270\textdegree rotations. This enhanced the total size and diversity of the dataset. Finally, two separate datasets were created with sample cutout sizes of 102$\times$102 pixels and 31$\times$31 pixels, respectively. They were used to train the CNN models of the two respective input sizes. Both datasets comprised 1628 samples for each class. Samples from the dataset for `extended-host', `point-host', and `hostless' classes are shown in Figure \ref{fig:transi_galaxy}, Figure \ref{fig:transi_stellar} and Figure \ref{fig:transi_hostless} respectively.

\subsection{An ensemble of CNN models}

The \texttt{NovaNet} module employs a logical combination of individual predictions from three CNN models to classify candidates. Depending on the cutout sizes of images transferred to the CNN input, the models are one of two types: (1) one model with an input shape of 31$\times$31 pixels, and (2) two models with an input shape of 102$\times$102 pixels.

\begin{figure*}
    \centering
    \begin{minipage}{0.49\textwidth}
        \centering
        \includegraphics[width=\textwidth]{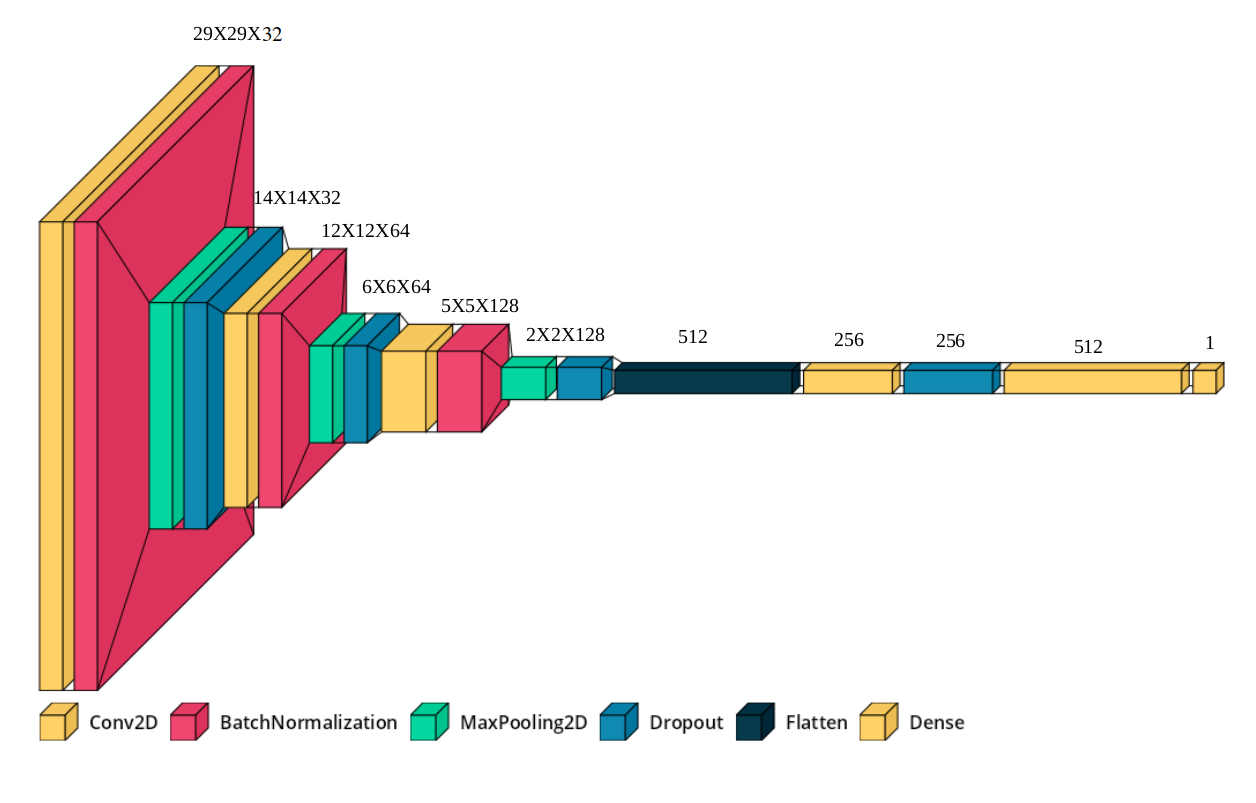}
        \caption{CNN architecture for the high-precision and high-recall real/bogus classifiers. The architecture diagram was generated using \texttt{visualkeras} (\protect\cite{Gavrikov2020VisualKeras}).}
        \label{fig:CNN_archi_63}
    \end{minipage}
    \hfill
    \begin{minipage}{0.4\textwidth}
        \centering
        \includegraphics[width=\textwidth]{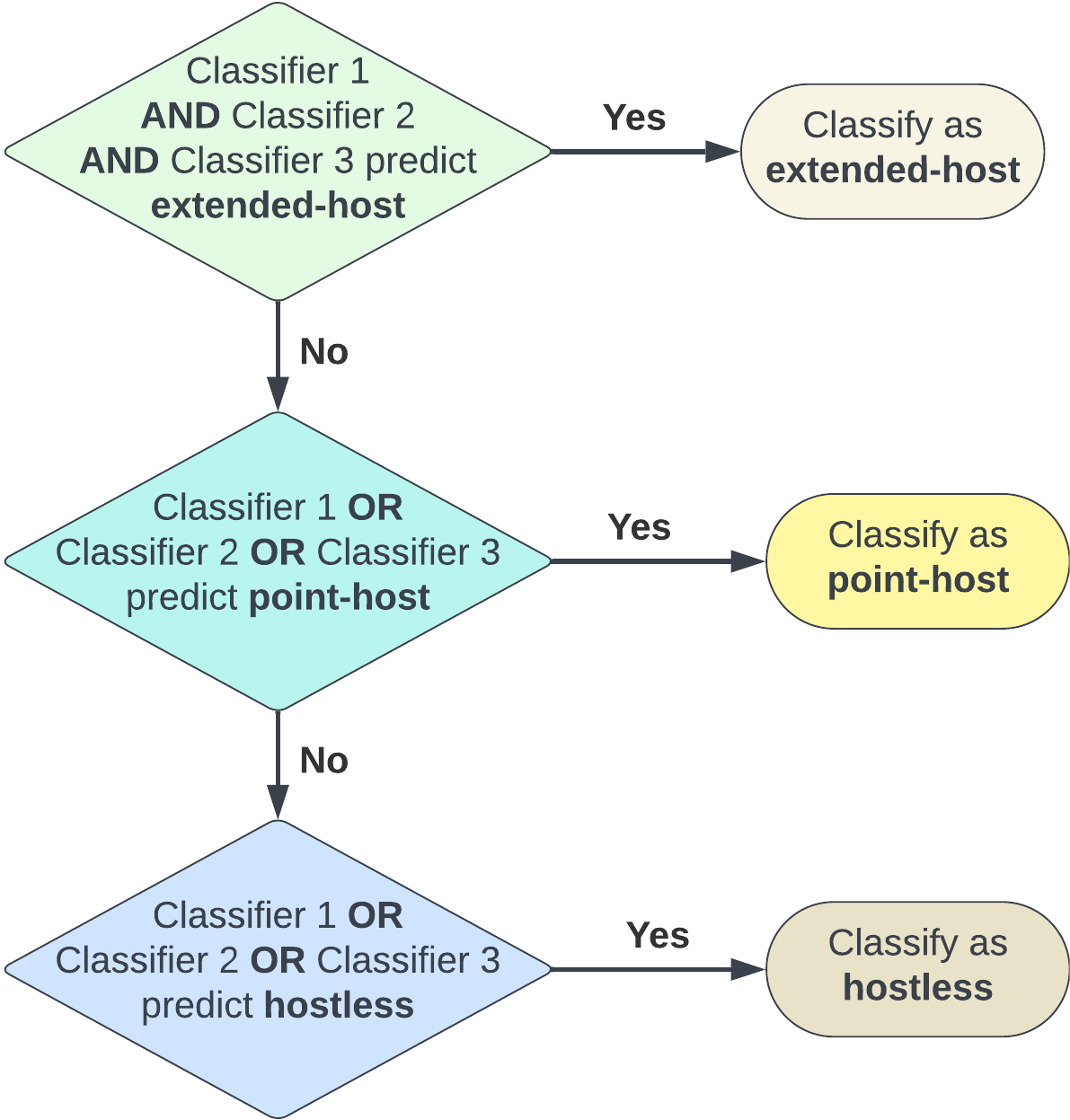}
        \caption{Logical scheme for the CNN transient candidate classifiers used in the \texttt{NovaNet} module. Classifier 1 has an input dimension of 31$\times$31 pixels, while classifiers 2 and 3 have input dimensions of 102$\times$102 pixels.}
        \label{fig:NovaNet_flowchart}
    \end{minipage}
\end{figure*}

The strategy of combining predictions from multiple classifiers proved effective in reducing false positive predictions for a specific class. For instance, to lower the false positive rate for `extended-host' objects and expedite the search for SNe candidates, the prediction criterion was set to require a positive detection from all three classifiers. The flowchart in Figure \ref{fig:NovaNet_flowchart} illustrates this logical scheme for combining model predictions. The three classes are positioned at different levels in the flowchart depending on class priority (i.e. `extended-host' objects are prioritised over `point-host', followed by `hostless'). 

\begin{figure*}
  \centering
  \begin{subfigure}{.49\linewidth}
    \includegraphics[width=\linewidth]{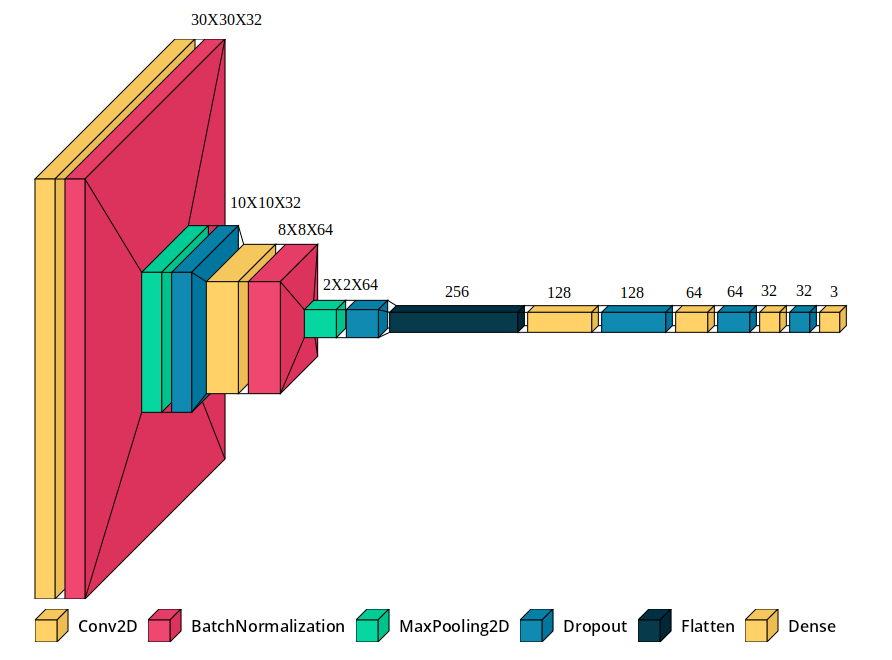}
    \caption{31$\times$31 pixel classifier}
    \label{fig:CNN_archi_novanet_31}
  \end{subfigure}
  \hfill
  \begin{subfigure}{.49\linewidth}
    \includegraphics[width=\linewidth]{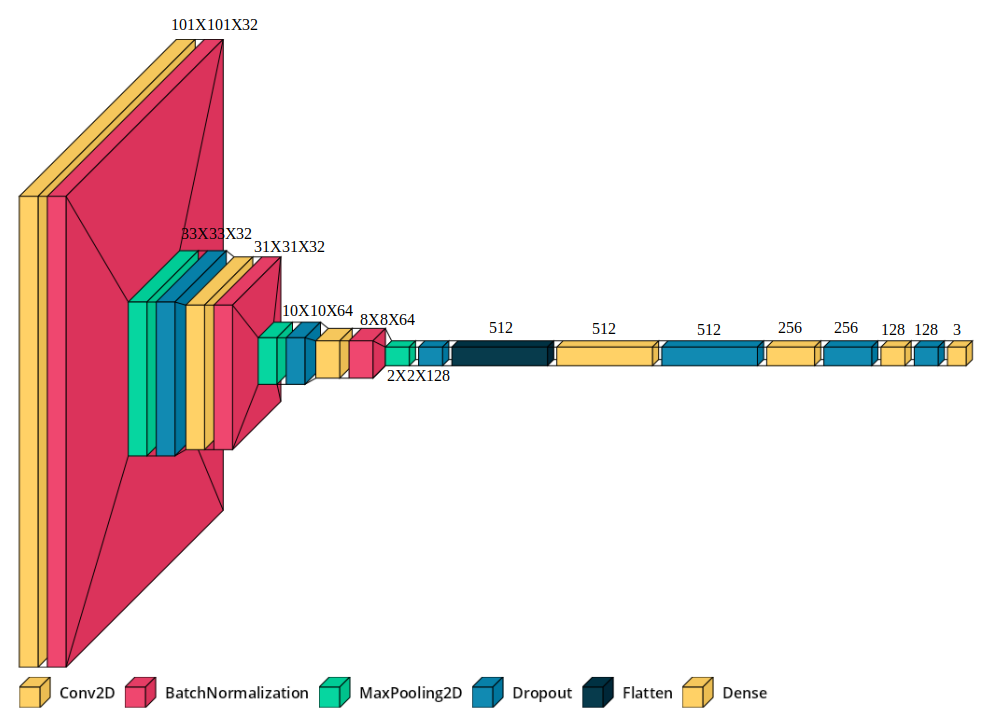}
    \caption{102$\times$102 pixel classifier}
    \label{fig:CNN_archi_novanet_102}
  \end{subfigure}
  \caption{Model architectures of CNN-based transient candidate classifier.}
  \label{fig:cnn_31_102_novanaet}
\end{figure*}

The 31$\times$31 pixel classifier and 102$\times$102 pixel classifier architectures have two and three convolutional layers in the feature extraction region, respectively. Each convolutional layer is coupled with a max-pooling layer of size 2$\times$2 or 3$\times$3. A combination of 2$\times$2 and 3$\times$3 filters are used. The fully connected region in both types of classifiers has 3 dense layers and an output layer with 3 dimensions, representing the 3 possible classes. The activation function used for all the layers except the output layer is \texttt{ReLU} and that for the output layer is \texttt{softmax}. Batch normalisation layers are also used with each convolutional layer. The CNN architectures of both classifier types are shown in Figure \ref{fig:cnn_31_102_novanaet}.
 
\subsection{Training of the CNN}\label{sec:Novanet_CNN}
For training the models, the loss function used was \texttt{categorical\_crossentropy} while the optimiser was \texttt{adam}. \texttt{L2} regularisation and \texttt{dropout} regularisation were used to prevent overfitting in both types of classifiers. The regularisation hyperparameter for the L2 regulariser was kept at 0.01. The dropout rate was 0.3 for the feature extraction/convolutional region and 0.5 for the fully connected region.

\begin{figure*}
  \centering
  \begin{tabular}{@{}c@{}}
    \includegraphics[width=.49\linewidth]{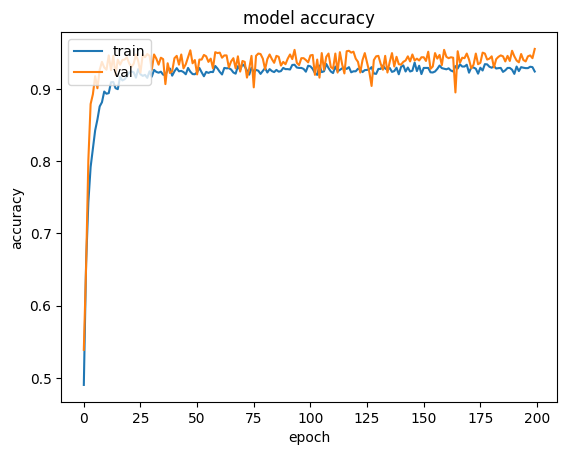} \\[\abovecaptionskip]
    % \small (a) accuracy curves
  \end{tabular}
  \hfill
  \begin{tabular}{@{}c@{}}
    \includegraphics[width=.49\linewidth]{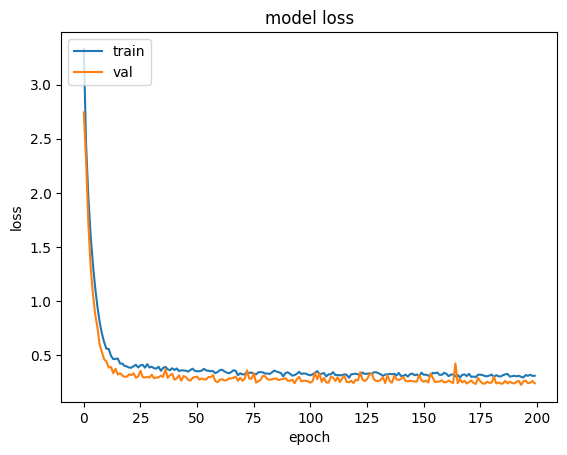} \\[\abovecaptionskip]
    % \small (b) loss curve
  \end{tabular}
  \caption{The accuracy and loss curves for the 31$\times$31 pixels CNN model for the transient candidate classifier.}
\label{fig:acc_loss_31_novanaet}
\end{figure*}

\begin{figure*}
  \centering
  \begin{tabular}{@{}c@{}}
    \includegraphics[width=.49\linewidth]{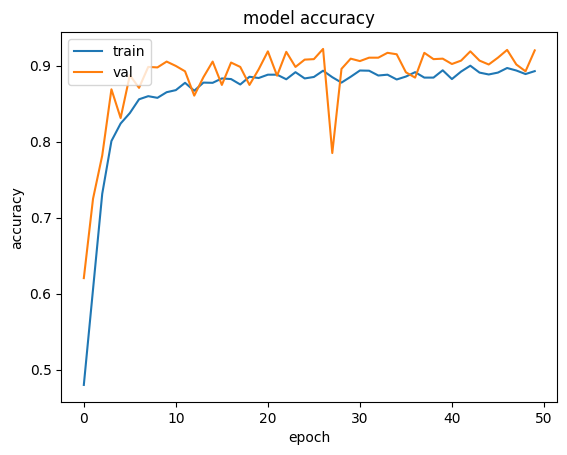} \\[\abovecaptionskip]
    % \small (a) accuracy curves
  \end{tabular}
  \hfill
  \begin{tabular}{@{}c@{}}
    \includegraphics[width=.49\linewidth]{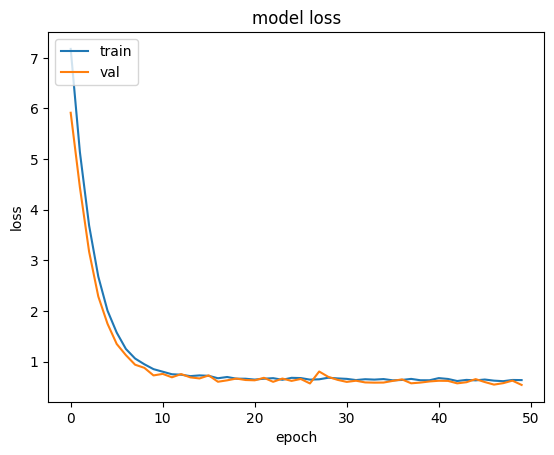} \\[\abovecaptionskip]
    % \small (b) loss curve
  \end{tabular}
  \caption{The accuracy and loss curves of a 102$\times$102 pixels CNN model trained for the transient candidate classifier.}
\label{fig:acc_loss_102_novanet}
\end{figure*}

Like the real/bogus classifiers, the samples from the dataset were standardised before training and the dataset was split into training and validation sets in an 80:20 ratio. Artificial noise was again added to the training samples for enhancing model sensitivity. The CNN architecture was written using Google's \texttt{TensorFlow} for \texttt{Python} as a \texttt{sequential} network. Further data augmentation was performed by applying \textit{rotation\_range}, \textit{height\_shift\_range} and \textit{width\_shift\_range} using \texttt{Keras} \texttt{ImageDataGenerator} module. The training was done using NVIDIA\textsuperscript{\tiny\textregistered} T4 GPU available with Google Colab\textsuperscript{\tiny\textregistered}. The 31$\times$31 pixels model was trained for 200 epochs while the two 102$\times$102 pixel models were trained for 50 epochs each. The final validation accuracy achieved was approximately 96\% for the 31$\times$31 pixel classifier and around 93\% and 95\% for the two 102$\times$102 pixel classifiers. The accuracy and loss curves for the 31$\times$31 pixel CNN classifier are shown in Figure \ref{fig:acc_loss_31_novanaet} and the same for one of the 102$\times$102 pixel CNN classifiers are shown in Figure \ref{fig:acc_loss_102_novanet}.   

\input{PyLMT_parameters}
%%%%%%%%%%%%%%%%%%%%%%%%%%%%%%%%%%%%%%%%%%%%%%%%%%

% Don't change these lines
\bsp	% typesetting comment
\label{lastpage}
\end{document}

%% file: ILMT_parameters.tex
\begin{table}
\centering
\caption{Telescope and detector parameters of the ILMT \protect\citep{BKumar2022}.}
\begin{tabular}{cc}
\hline
Parameter & Value \\
\hline
Aperture size & 4.0-m diameter \\
f-ratio &  $\sim$2.4 \\ 
Field of view & 22$'$$\times$22$'$ \\
Accessible sky area &  $\sim$36 degree\textsuperscript{2} per night \\
Bowl rotation period & 8.02 sec \\
CCD Size & 4096$\times$4096 pixels \\
Pixel size & 0$''$.328 pixel\textsuperscript{-1} \\
Readout noise & 5.0 e$^{-}$ \\
Gain & 4.0 e$^{-}$/ADU \\
Integration time & 102.36 sec \\
filters & SDSS \textit{g}$'$, \textit{r}$'$, \textit{i}$'$ \\
\hline
\end{tabular}
\label{tab:ILMT_parameters}
\end{table}

%% file: transisearch_threshold.tex
\begin{table*}
    \centering
    \caption{Additional threshold parameters and their allowed range for candidate sources detected in the difference images by CNN-based real/bogus classifier in the \texttt{TransiSearch} module.}
    \begin{tabular}{P{0.2\linewidth}P{0.45\linewidth}P{0.2\linewidth}}
    \hline
      Parameter (units) & Description & Range\\
      \hline\\
       amp\_s (ADUs)  & Amplitude of 2D Gaussian profile fitted to the source in science image corresponding to detection in difference image & < 30000\\\\
      amp\_r (ADUs) & Amplitude of 2D Gaussian profile fitted to the source in reference image corresponding to a detection in difference image & < 30000\\\\
      sigma\_c\_x (pixels)&FWHM along x-axis of candidate detected in difference image&4.7 -- 14.1\\\\
      sigma\_c\_y (pixels)&FWHM along y-axis of candidate detected in difference image&4.7 -- 14.1\\\\
      sigma\_c\_x\_45 (pixels)&FWHM along x-axis of candidate source detected in difference image and rotated by 45\textdegree &4.7 -- 14.1\\\\
      sigma\_c\_y\_45 (pixels)&FWHM along y-axis of candidate source detected in difference image and rotated by 45\textdegree &4.7 -- 14.1\\\\
      % sigma\_c\_x\_315 (pixels)&FWHM along x-axis of candidate source detected in difference image and rotated by 315\textdegree &4.7 -- 14.1\\\\
      % sigma\_c\_y\_315 (pixels)&FWHM along y-axis of candidate source detected in difference image and rotated by 315\textdegree &4.7 -- 14.1\\\\
      sigma\_c\_x/sigma\_c\_y (ratio)&ratio of FWHMs of the candidate source along x-axis and y-axis&0.5 -- 2.0\\\\
      flux\_c (ADUs)&Total counts of the candidate source within aperture of predefined radius of 10 pixels&> 1000\\\\
      \hline
    \end{tabular}
    \label{tab:transisearch_threshold}
\end{table*}

%% file: candidate_classifier_dataset.tex
\begin{table*}
\centering
\captionof{table}{Summary of training datasets, pre-processing and augmentations applied for 31$\times$31 pixel and 102$\times$102 pixel transient candidate classifiers.}
\begin{tabular}{|l|c|c|c|}
\hline
\textbf{Candidate class} & \textbf{Number of samples} & \textbf{Data pre-processing} & \textbf{Data augmentation}\\
\hline
\textbf{Extended-host}  & 1628 & \multirow{3}{*}{standardisation} & \multirow{3}{*}{rotations, height shift, width shift} \\
\textbf{Point-host}  & 1628  &&     \\
\textbf{Hostless}  & 1628 &&\\        
\hline
\end{tabular}
\label{table:sample_counts_tr}
\end{table*}

%% file: real_bogus_dataset.tex
\begin{table*}
\centering
\captionof{table}{Summary of training datasets, pre-processing and augmentations applied for high-precision and high-recall real/bogus classifiers.}
\begin{tabular}{|l|c|c|c|p{5cm}|}
\hline
        \textbf{Dataset type} & \textbf{Real samples} & \textbf{Bogus samples} & \textbf{Data pre-processing} & \textbf{Data augmentation}\\
\hline
\textbf{High-recall dataset}  & 2705       & 2527  & standardisation & rotations, random horizontal flips\\
\textbf{High-precision dataset} & 1000       & 2527  & standardisation & rotations, random horizontal flips, height shifts, width shifts \\
\hline
\end{tabular}
\label{table:sample_counts_rb}
\end{table*}

%% file: novanet.tex
\begin{table*}
\centering
\captionof{table}{Classifications made by the \texttt{NovaNet} transient candidate classifier alongside class-wise true positive predictions. Classifications were compiled corresponding to the real sources correctly identified in the ILMT subtracted images from the observation cycle of October-November 2022.}
\begin{tabular}{|m{2cm}|m{2cm}m{2cm}m{2cm}|m{2cm}m{2cm}m{2cm}|}
    \hline
    \textbf{LST Field} & predicted extended-host & predicted point-host & predicted hostless & extended-host true positives & point-host true positives & hostless true positives \\
    \hline
     0h 42m & 0 & 3 & 37 & 0 & 3 & 36 \\
     1h 8m & 1 & 4 & 75 & 0 & 3 & 75 \\
     1h 26m & 2 & 5 & 73 & 1 & 5 & 73 \\
     1h 45m & 4 & 7 & 99 & 2 & 3 & 99 \\     
     2h 02m & 0 & 7 & 120 & 0 & 7 & 118 \\
     2h 20m & 2 & 6 & 149 & 0 & 4 & 149 \\
     4h 11m & 6 & 30 & 200 & 0 & 27 & 200 \\
     4h 50m & 0 & 44 & 158 & 0 & 40 & 156 \\
     5h 07m & 1 & 58 & 199 & 0 & 52 & 198 \\
     6h 37m & 3 & 41 & 180 & 1 & 41 & 180 \\
     \hline
\end{tabular}
\hspace{3cm}
\label{tab:novanet}
\end{table*}

%% file: ILMTDiff_threshold.tex
\begin{table*}
    \centering
    \caption{Allowed range of parameters for a source in science and reference image for it to be considered for computation of convolutional kernel by the \texttt{ILMTDiff} module.}
    \begin{tabular}{P{0.2\linewidth}P{0.45\linewidth}P{0.2\linewidth}}
    \hline
      Parameter (units) & Description & Range\\
      \hline\\
      peak\_thresh (ADUs)  & Peak amplitude of source detected in reference image & 2500 -- 55000 (1500 -- 55000 in case of insufficient detection)\\\\
      sigma\_x\_s (pixels)&FWHM along x-axis for a source in science image&3.5 -- 16.5\\\\
      sigma\_y\_s (pixels)&FWHM along y-axis for a source in science image&3.5 -- 16.5\\\\
      sigma\_x\_s/sigma\_y\_s (ratio)&ratio of FWHMs of the sources in science image along x-axis and y-axis&0.5 -- 2.0\\\\
      sigma\_x\_r (pixels)&FWHM along x-axis for a source in reference image&3.5 -- 16.5\\\\
      sigma\_y\_r (pixels)&FWHM along y-axis for a source in reference image&3.5 -- 16.5\\\\
      sigma\_x\_r/sigma\_y\_r (ratio)&ratio of FWHMs of the sources in reference image along x-axis and y-axis&0.5 -- 2.0\\\\
      \hline
    \end{tabular}
    \label{tab:ILMTDiff_threshold}
\end{table*}

%% file: PyLMT_parameters.tex
\begin{table*}
\centering
\caption{Tunable parameters of the PyLMT transient detection pipeline.}
\begin{tabular}{|m{5cm}|m{4cm}|m{7cm}|}
    \hline
    \textbf{Parameter} & \textbf{Default value} & \textbf{Use}\\
    \hline
    alignment\_sigma & 5 & Standard deviation above which a source should be considered for aligning the images by 3-point asterisms using \texttt{astroalign}\\
    \hline
    realign & False & re-aligning of the science and reference cutouts passed to the \texttt{ILMTDiff} algorithm using \texttt{astroalign}. It defaults to False if primary alignment is being performed using WCS information\\
    \hline
    wcs\_align & True & Boolean flag to align the science and reference images using WCS information\\
    \hline
    background\_grid\_size & (512,512) & Grid size in pixels used in performing spatially varying median background subtraction\\
    \hline
    diff\_method & `Bramich' & Parameter to choose the method for image subtraction. Alard-Lupton method can be selected with the `Alard-Lupton' option\\    
     \hline
    extend\_size & 0 & Parameter to increase the size of the cropped image along RA axis for subtraction in case of sparse field. New size of image along RA becomes (2$\times$extend\_size+1)$\times$1024 pixels\\      
     \hline
    stamp\_size & 21 & size in pixels of square cutouts of sources extracted from science and reference images for kernel optimisation during image subtraction\\      
     \hline
     selection\_thresh & (30000,30000,2,6) & Bounds on fitted 2D Gaussian parameters for detected candidates. It represents the following parameters in sequence (fitted Gaussian amplitude of science image source, fitted Gaussian amplitude of reference image source, minimum sigma of Gaussian fitted to candidate, maximum sigma of Gaussian fitted to candidate)\\    
     \hline
    detection\_threshold & 5 & Detection threshold for source detection in terms of standard deviations above noise\\  
    \hline
    classification\_threshold & 0.5 & Classification threshold for the real/bogus classifier\\ 
     \hline
    overlap\_width & 26 & overlap width in pixel scale for two neighbouring images being cropped for subtraction\\    
     \hline     
    show\_PyLMT & True & Boolean flag for printing results for every subtracted chunk of a full-frame image\\    
     \hline 
    show\_stellar & True & Boolean flag for printing detected variable star candidates\\    
     \hline 
    show\_asteroids & True & Boolean flag for printing detected asteroid candidates\\    
     \hline 
    show\_extragalactic & True & Boolean flag for printing detected extra-galactic candidates\\ 
     \hline
    advanced\_filtering & True & Boolean flag to prepare a filtered list of alerts after rejecting cataloged asteroids and possible variable stars\\
    \hline
    peak\_thresh\_ & 2500 & detection threshold (in counts) to select sources for kernel optimisation while performing image subtraction in \texttt{ILMTDiff}\\
    \hline
    delta\_peak\_thresh & 1000 & the quantity by which the value of the \textit{peak\_thresh\_} parameter is lowered successively for up to two iterations if no sources were found at the default threshold\\
    \hline
    band & `i' & Spectral band in which the image was acquired\\    
     \hline
     sensitive\_mode & False & Boolean flag for enabling median filtering of input images to real/bogus classifier. This helps detect faint transients but also increases the number of false positive detections \\
    \hline
    ILMT\_ramping\_exclude & True & Boolean flag for enabling removal of TDI ramping region in ILMT images. If no ramping region is present then the image remains unaltered. This assumes that the image is of size 4096$\times$40960 pixels or 4096$\times$36864 pixels\\
    \hline
     over\_detection\_exit & False & Boolean flag to enable auto-termination of transient detection for a particular pair of science and reference images in case of too many detections (likely caused due to failed subtraction)\\
    \hline
     print\_checkpoints & False & Boolean flag to print the status of data processing for diagnostic purposes\\
    \hline
\end{tabular}
\label{tab:PyLMT_parameters}
\end{table*}